\documentclass[11pt]{article}

\usepackage[utf8]{inputenc}
\usepackage[T1]{fontenc}
\usepackage{lmodern}
\usepackage[margin=1in]{geometry}
\usepackage{amsmath,amssymb,amsthm}
\usepackage{bm}
\usepackage{graphicx}
\usepackage{booktabs}
\usepackage{multirow}
\usepackage{array}
\usepackage{float}
\usepackage{caption}
\usepackage{subcaption}
\usepackage{xcolor}
\usepackage{siunitx}
\usepackage{enumitem}
\usepackage{CJKutf8}
\usepackage[hidelinks]{hyperref}
\usepackage{cleveref}

\graphicspath{{Images/}}

\title{\textbf{Multimodal Three-Class Alzheimer’s Disease Classification: The MCI Bottleneck}}

\author{
Lorenzo Tanzi\thanks{MOX Laboratory, Department of Mathematics, Politecnico di Milano, Milan, Italy. Corresponding author.}
\and
Chunfeng Lian\thanks{School of Mathematics and Statistics, Xi'an Jiaotong University, Xi'an, China.}
\and
Lara Cavinato\thanks{MOX Laboratory, Department of Mathematics, Politecnico di Milano, Milan, Italy.}
\and
for the Alzheimer's Disease Neuroimaging Initiative\thanks{Data used in preparation of this article were obtained from the Alzheimer's Disease Neuroimaging Initiative (ADNI) database (\href{https://adni.loni.usc.edu}{adni.loni.usc.edu}). As such, the investigators within the ADNI contributed to the design and implementation of ADNI and/or provided data but did not participate in analysis or writing of this report. A complete listing of ADNI investigators can be found at: \url{http://adni.loni.usc.edu/wp-content/uploads/how_to_apply/ADNI_Acknowledgement_List.pdf}}
}

\date{
\small
\texttt{lorenzo.tanzi@mail.polimi.it} \qquad
\texttt{chunfeng.lian@xjtu.edu.cn} \qquad
\texttt{lara.cavinato@polimi.it}
}
\begin{document}
\maketitle

\begin{abstract}
Alzheimer's disease is clinically staged as a three-class progression: cognitively normal (CN), mild cognitive impairment (MCI) and Alzheimer's disease (AD). The intermediate MCI class is biologically heterogeneous and overlaps with both extremes and it is the dominant source of classification error. This work develops a leakage-controlled framework for subject-level CN/MCI/AD classification that fuses four sources of information: a 3D Tau PET model, a 3D structural MRI model, a tabular ROI-and-plasma model and a hierarchical PET--plasma model. All branches are derived from a custom preprocessing and feature-extraction pipeline applied to $881$ aligned subjects from the ADNI cohort. They are aligned at the subject level and evaluated under identical, no-leak cross-validation folds, so that the multimodal fusion, built entirely on synchronized out-of-fold predictions, avoids optimistic bias. The final model is a fixed, parameter-free convex combination of the four branches. A nested superlearner serves as a robustness analysis and reproduces, rather than improves on, the fixed weights. The fusion improves significantly over every individual branch, with the largest gain on the MCI class, while a logistic meta-learner provides a complementary, MCI-oriented operating point on the same accuracy-sensitivity frontier. The work is presented as an honest baseline. It shows that multimodal fusion improves robustness, but that MCI remains the central bottleneck. This reflects a limit of the cross-sectional information rather than of the fusion mechanism and motivates reformulating the task as a continuous progression problem.

\end{abstract}

\section{Introduction}

Alzheimer's disease (AD) is a progressive neurodegenerative disorder and the leading cause of dementia in older individuals~\cite{monfared2022epidemiology}. Its worldwide burden grows as the population ages and is widely believed to be underestimated due to systematic underdiagnosis~\cite{monfared2022epidemiology, weidner2023worldalzheimerreport}. Its two neuropathological hallmarks, extracellular amyloid-$\beta$ plaques and intracellular neurofibrillary tau tangles, accumulate years before any visible cognitive symptom appears, which makes early and accurate diagnosis difficult and valuable~\cite{jack2018niaaa, cody2024tauamyloidtimeline}.

Clinically, the disease is described as a progression along a continuum and defined as a three-stage problem: cognitively normal (CN), mild cognitive impairment (MCI) and Alzheimer's disease (AD). MCI denotes the intermediate state in which a measurable cognitive decline can be detected but everyday functioning is preserved~\cite{lee2023mciethics}.  A substantial and variable fraction of MCI subjects progress to AD dementia, with reported conversion ranging from roughly $40\%$ to $75\%$ depending on the population and whether the diagnosis is purely clinical or biomarker-supported~\cite{monfared2022epidemiology}. This is why the MCI class carries most of the clinical interest, because it is where intervention is still plausible but the prognosis is most uncertain.

The two endpoints of the spectrum are relatively easy to separate, since CN and AD differ markedly in both neurobiological and cognitive terms. The difficulty concentrates in the MCI group, whose definition spans a biologically heterogeneous population~\cite{lee2023mciethics} that overlaps, at the level of imaging and biomarker features, with both other classes~\cite{aghdam2025mladreview}. This overlap motivates richer, multi-source representations of the disease state. \\

Neuroimaging and biological biomarkers offer complementary windows onto the Alzheimer's pathological cascade. The National Institute on Aging and Alzheimer's Association (NIA-AA) research framework organises them according to the ATN system: amyloid (A), pathologic tau (T) and neurodegeneration (N), each capturing a distinct and only partially dependent dimension of the disease process~\cite{jack2018niaaa}. The modalities used in this work provide complementary measures of these dimensions. Tau PET quantifies the spatial deposition of neurofibrillary tangles, whose regional burden tracks cognitive decline and stage (the T axis)~\cite{cody2024tauamyloidtimeline, iqbal2024tauad}; structural MRI measures macroscopic neurodegeneration as atrophy (the N axis), downstream of and complementary to molecular imaging~\cite{jack2018niaaa}; plasma biomarkers, in particular phosphorylated-tau species, provide minimally invasive systemic proxies of pathology with performance approaching that of cerebrospinal-fluid and imaging markers~\cite{leuzy2022bloodbiomarkers, arslan2024bloodbiomarkers}; and region-of-interest (ROI) features extracted from the normalised images give a
compact, interpretable summary of regional intensity.

Each source captures a different and only partly overlapping face of the disease and none is expected to resolve the three-class task on its own, least of all in the ambiguous MCI group. This complementarity is the premise on which the work is built: multimodal fusion strategies integrate heterogeneous data at the subject level to exploit it~\cite{dwivedi2022multimodalfusionad, wang2024crossmodalad}. All data used in this work are taken from the ADNI cohort~\cite{petersen2010adni} and processed through a reproducible preprocessing and feature-extraction pipeline developed for this work.\\

Multimodal machine learning for Alzheimer's staging has been studied
extensively~\cite{aghdam2025mladreview, dwivedi2022multimodalfusionad}, yet several methodological issues recur and remain insufficiently addressed.

First, the simultaneous CN/MCI/AD classification task is uncommon despite the large number of research and publications on this topic. A major part of the literature reduces the problem to a set of easier binary comparisons, CN-versus-AD in particular, where the intermediate class in particular is often ignored~\cite{aghdam2025mladreview}. A further part operates on selected 2D slices rather than on the whole volume, relying on transfer learning from natural-image networks and on heavy slice-level preprocessing rather than on a true volumetric model. Where very high three-class accuracies are reported, they are frequently accompanied by signs of optimistic evaluation and poor out-of-sample generalisation~\cite{aghdam2025mladreview}. A leakage-controlled, subject-level, full-volume treatment of
the three-class task therefore remains less established than it appears.

Second, multimodal fusion is often performed without adequate control of information leakage. When the predictions feeding a fusion model are generated on the same data used to train the base models, the resulting estimates are optimistically biased. A solid stacking procedure instead requires that every fusion input is an out-of-fold (OOF) prediction, so that the meta-learner never sees an in-sample prediction from any branch it combines. This is well explained in the principle of stacked generalization~\cite{wolpert1990stackedgeneralization} and of its cross-validated formalisation, the super learner~\cite{vanderlaan2007superlearner}.

Third, combining several modalities requires that all branches are aligned at the subject level and evaluated under identical cross-validation folds. Without such synchronization, differences in subject matching or fold assignment across branches confound both the single-branch comparison and the fusion~\cite{aghdam2025mladreview}. This work addresses these gaps jointly, through a leakage-controlled, subject-level fusion framework built entirely on synchronized out-of-fold predictions.\\

The objective of this work is to establish a rigorous and reproducible baseline for multimodal CN/MCI/AD classification, relying on well established model architectures in literature. Its emphasis is on a clean, leakage-controlled evaluation and on a thorough account of what multimodal fusion can and cannot deliver on the intermediate class. Its main objectives are:
to build a complete, reproducible preprocessing and ROI-extraction pipeline for MRI and Tau PET data from the ADNI
cohort~\cite{petersen2010adni, aisen2024adniclinicalcore}; to compare image-based and tabular models for subject-level CN/MCI/AD classification on a common, synchronized cohort; to assess the diagnostic contribution of plasma biomarkers and Harvard--Oxford ROI features; to design a leakage-controlled multimodal
fusion built on synchronized out-of-fold predictions from all branches; and, lastly, to analyse the complementarity of the modalities and the persistent difficulty of the MCI class.

The main contributions are:
\begin{itemize}
    \item an MRI-centric PET preprocessing workflow, performing brain extraction with SynthStrip~\cite{hoopes2022synthstrip}, bias-field correction, affine registration of the structural MRI to MNI152 space and a single-interpolation PET-to-MNI transform obtained by composing the rigid PET-to-MRI and affine MRI-to-MNI maps, which avoids accumulating interpolation error;
    \item a robust, generalizable Harvard-Oxford ROI extraction procedure applied to spatially normalised Tau PET volumes;
    \item a subject-level comparison of four branches, a 3D Tau PET model, a 3D structural MRI model, a tabular ROI-and-plasma model and a hierarchical PET-plasma model, implemented within a deep-learning framework for medical imaging~\cite{cardoso2022monai};
    \item a leakage-controlled multimodal fusion in which a fixed, transparent convex combination is the primary model and a nested convex superlearner~\cite{vanderlaan2007superlearner} serves as a robustness analysis that confirms, rather than improves, the fixed choice.\\
\end{itemize}

The remainder of this paper is organised as follows. Section~\ref{sec:background} provides the clinical and methodological background. Section~\ref{sec:dataset} describes the ADNI data, the preprocessing pipeline and the feature extraction. Section~\ref{sec:methods} presents the four branches and the fusion methodology. Section~\ref{sec:results} reports the single-branch and fusion results on the synchronized cohort, with implementation details deferred to the appendices. Section~\ref{sec:discussion} discusses the findings, their limitations and the directions they open and Section~\ref{sec:conclusion}
concludes.
\section{Background}
\label{sec:background}

\subsection{Alzheimer's disease and clinical staging}

Alzheimer's disease is conventionally described as a progression along a clinical continuum, from a cognitively normal (CN) state, through mild cognitive impairment (MCI), to AD dementia~\cite{monfared2022epidemiology}. CN
subjects show no clinically significant problems; AD dementia involves cognitive impairment severe enough to compromise daily functioning; MCI occupies the intermediate position, with measurable decline but largely preserved everyday
functioning~\cite{lee2023mciethics}. The NIA-AA research framework has shifted the definition of AD towards a
biological one, identifying the disease by its underlying amyloid and tau pathology rather than by symptoms alone~\cite{jack2018niaaa}. This has reinforced
the value of early detection, since the pathology accumulates years before symptoms appear~\cite{cody2024tauamyloidtimeline}.

The principal difficulty in three-class staging is the MCI group, which is clinically and biologically heterogeneous~\cite{lee2023mciethics}. In particular, some individuals with MCI remain clinically stable over several years, whereas others progress to AD. Distinguishing these two trajectories remains difficult with current screening methods~\cite{aghdam2025mladreview}. This heterogeneity, together with the overlap of MCI with both CN and AD at the feature level, is why MCI is consistently the hardest class to recognise and the dominant source of error in three-class classification.

\subsection{Imaging and biomarker modalities}

Within the ATN framework introduced above, the modalities considered here explore distinct axes of the disease and their respective strengths and limitations motivate combining them rather than relying on any single one.

\paragraph{Tau PET.}
Tau positron emission tomography uses radiotracers that bind aggregated tau, enabling the detection of neurofibrillary tau pathology, one of the two defining lesions of AD corresponding to the T component in the framework~\cite{iqbal2024tauad, jack2018niaaa}. Tau accumulation follows a relatively standard spatial pattern beginning in medial temporal regions and later spreading to other cortical areas~\cite{cody2024tauamyloidtimeline}; this relationship between tau topography and stage makes Tau PET informative for distinguishing disease stages.

\paragraph{Structural MRI.}
Structural MRI non-invasively measures macroscopic brain morphology and quantifies neurodegeneration as cortical thinning and regional atrophy, in particular in the medial temporal lobe and hippocampus (the N axis)~\cite{jack2018niaaa}. However atrophy is not specific to AD, it overlaps with normal ageing and with other neurodegenerative conditions. Moreover, the CN/early-MCI morphological difference is subtle and MRI-based models are sensitive to scanner, protocol and cohort differences~\cite{aghdam2025mladreview, abbasi2024mriharmonization}.

\paragraph{Plasma biomarkers.}
Plasma biomarkers, in particular phosphorylated-tau species, are minimally invasive indicators of AD pathology. In selected settings, their diagnostic performance can approach that of established cerebrospinal-fluid and imaging markers, while amyloid-related plasma ratios add complementary information on amyloid status~\cite{leuzy2022bloodbiomarkers, arslan2024bloodbiomarkers}. Since they reflect systemic correlates of pathology rather than its spatial distribution, they can be regarded as complementary to the regional information of Tau PET and MRI~\cite{leuzy2022bloodbiomarkers} and their low cost motivates their inclusion as a tabular modality.

\paragraph{ROI-based representations.}
Rather than processing whole volumes, each image can be summarised into region-of-interest (ROI) features defined by a standard anatomical atlas; in this work the Harvard--Oxford atlas is applied to the normalised MRI and Tau PET volumes. ROI representations drastically reduce dimensionality and are directly interpretable in terms of named anatomical structures, a property largely lost in end-to-end voxel models~\cite{aghdam2025mladreview}. They thus provide a strong, reproducible baseline and a natural complement to the image-based branches.

\subsection{Background on modelling and multimodal fusion}
Supervised classification of neuroimaging data for AD has been approached along two complementary lines: traditional models trained on hand-crafted features and deep models trained end-to-end on images~\cite{aghdam2025mladreview}. The two families offer different trade-offs. Deep image models, such as 3D convolutional networks, can learn discriminative spatial patterns directly from the volumes, but they require large training sets and are sensitive to acquisition protocol and cohort variability~\cite{sahumbaiev2018hadnet, abbasi2024mriharmonization}. Tabular models on ROI and biomarker features are more compact, interpretable and often more robust in the moderate-sample regimes typical of multimodal neuroimaging studies, at the cost of spatial detail. Neither family dominates the other, and the literature suggests that a single modality is rarely sufficient for the three-class task.

Since different modalities capture complementary aspects of the disease, combining them is a natural way to improve robustness~\cite{dwivedi2022multimodalfusionad, wang2024crossmodalad}. The literature distinguishes broadly between early fusion, where features from all modalities are concatenated before a single model, and late fusion, where each modality is modelled independently and the resulting predictions are combined. Late fusion is attractive when the modalities are heterogeneous (3D volumes, tabular biomarkers) and when individual branches are to be trained and evaluated separately. A classic late-fusion strategy is stacking, in which a meta-learner is trained on the outputs of the base models~\cite{wolpert1990stackedgeneralization}. The super learner formalises this with cross-validation to construct a leakage-controlled ensemble~\cite{vanderlaan2007superlearner}.

The validity of stacking, however, depends on how the base predictions are generated. If the meta-learner is trained on predictions the base models made for their own training data, the inputs are optimistically biased and the fusion is inflated. Avoiding this requires that all fusion inputs be out-of-fold predictions, produced for each subject only by models that never saw that subject in training. The risk of such leakage and of over-optimistic evaluation from improper cross-validation is well documented~\cite{cawley2010overfitting}, and the limited generalizability of some published AD models reinforces the need for a strictly leakage-controlled protocol~\cite{aghdam2025mladreview}.
\section{Dataset and Preprocessing}
\label{sec:dataset}

\subsection{Dataset overview}

Data used in the preparation of this article were obtained from the Alzheimer's Disease Neuroimaging Initiative (ADNI) database (adni.loni.usc.edu). The ADNI was launched in 2003 as a public-private partnership, led by Principal Investigator Michael W. Weiner, MD. The primary goal of ADNI has been to test whether serial magnetic resonance imaging (MRI), positron emission tomography (PET), other biological markers, and clinical and neuropsychological assessment can be combined to measure the progression of mild cognitive impairment (MCI) and early Alzheimer's disease (AD). For up-to-date information, see \url{www.adni-info.org}.

For each subject, several data modalities are present: structural T1-weighted MRI, tau positron emission tomography (Tau PET, tracer AV1451), together with plasma biomarkers and demographic variables (age, sex). From the spatially normalised images, region-of-interest (ROI) features are additionally extracted, creating a compact tabular representation that complements the image-based modalities.

Because multimodal analysis requires complete subject-level alignment across all branches, the final study cohort comprises $881$ subjects with synchronized out-of-fold predictions from Tau PET, structural T1-weighted MRI, ROI-based features and plasma biomarkers. This cohort is used consistently for all standalone branch comparisons and for the final multimodal fusion. The class distribution is imbalanced, with $512$ CN subjects, $272$ MCI subjects and $97$ AD subjects; CN is therefore the majority class and AD the minority class. This imbalance directly motivates the class-balanced training and evaluation strategies described in Section~\ref{sec:methods}.

\subsection{Subject matching and synchronized folds}

Multimodal fusion requires that the predictions of all branches refer to the same subjects and be evaluated under an identical partitioning of the data. Two design choices enforce this consistency.

First, all analyses are performed at the subject level. When a subject contributes more than one scan, predictions are aggregated to a single subject-level prediction and the train/test partition is defined over subjects rather than scans. This prevents scans from the same subject appearing simultaneously in training and test sets, a common source of leakage in neuroimaging studies~\cite{cawley2010overfitting}.

Second, all branches share a single, fixed set of synchronized cross-validation folds, defined once at the subject level using stratified group $k$-fold partitioning, which preserves the class proportions across folds while keeping all scans of a subject within the same fold. Concretely, a fold assignment is computed for the aligned cohort and stored as a fixed table and every branch (PET, MRI, ROI/plasma and the hierarchical model) is trained and evaluated against exactly the same five folds. This guarantees that, for any given subject, the out-of-fold predictions produced by the different branches are mutually comparable and can be safely combined by the fusion model.

The reason for restricting the final multimodal comparison to the aligned $881$-subject cohort is precisely this requirement of synchronization. Only for these subjects are all branches defined and all out-of-fold predictions available under the same folds. Evaluating the fusion on this cohort therefore ensures a fair, leakage-controlled comparison in which every model is assessed on exactly the same subjects and partitions.

\subsection{MRI preprocessing}

Structural T1-weighted MRI volumes are processed through a fixed, subject-independent pipeline that produces, for each scan, a brain-extracted volume in standard MNI space together with the corresponding brain mask and the spatial transformation to MNI.

Each volume is first conformed to a standard orientation and resolution ($1\,\text{mm}$ isotropic), providing a consistent geometric reference for the subsequent steps. Brain extraction is then performed with SynthStrip, a learning-based skull-stripping method that is robust across image types and acquisition protocols~\cite{hoopes2022synthstrip} and a binary brain mask is derived from the brain-extracted volume. Intensity inhomogeneity (bias field) is corrected using the FSL FAST method, based on a hidden Markov random field model and the expectation-maximization algorithm~\cite{zhang2001hmmemsegmentation}. The bias-corrected brain volume is then registered to the $1\,\text{mm}$ MNI152 template with an affine (12-degrees-of-freedom) linear registration~\cite{jenkinson2002flirt} and the estimated transformation is applied to the brain mask using nearest-neighbour interpolation, propagating it into MNI space.

For each MRI, the pipeline therefore stores the brain-extracted volume in MNI space, the corresponding MNI brain mask, the native conformed image and the estimated image-to-MNI transformation matrix. The transformation matrix is retained explicitly because it is reused, without recomputation, in the PET pipeline described below.

\subsection{PET preprocessing}

PET volumes are aligned to the same MNI space as the MRI by reusing the MRI-derived transformation, so that all modalities share a common anatomical reference. When a PET acquisition is provided as a dynamic (4D) volume, it is first reduced to a static 3D image by temporal averaging (static volumes are used directly).

The static PET image is then rigidly registered (6 degrees of freedom) to the corresponding brain-extracted MRI of the same subject~\cite{jenkinson2002flirt}, providing a PET-to-MRI transformation. Rather than warping PET to MNI in two separate interpolation steps, the PET-to-MRI and MRI-to-MNI transformations are composed into a single PET-to-MNI transformation,
\begin{equation}
    T_{\text{PET}\rightarrow\text{MNI}} = T_{\text{MRI}\rightarrow\text{MNI}} \circ T_{\text{PET}\rightarrow\text{MRI}},
    \label{eq:pet_compose}
\end{equation}
which is applied to the original PET volume in a single resampling operation. Composing the transforms in this way avoids the accumulation of interpolation errors that would arise from two consecutive resampling steps. Finally, the PET volume in MNI space is skull-stripped using the MRI-derived MNI brain mask, ensuring that PET and MRI share exactly the same brain support.

\begin{figure}[H]
    \centering
    \includegraphics[width=\textwidth]{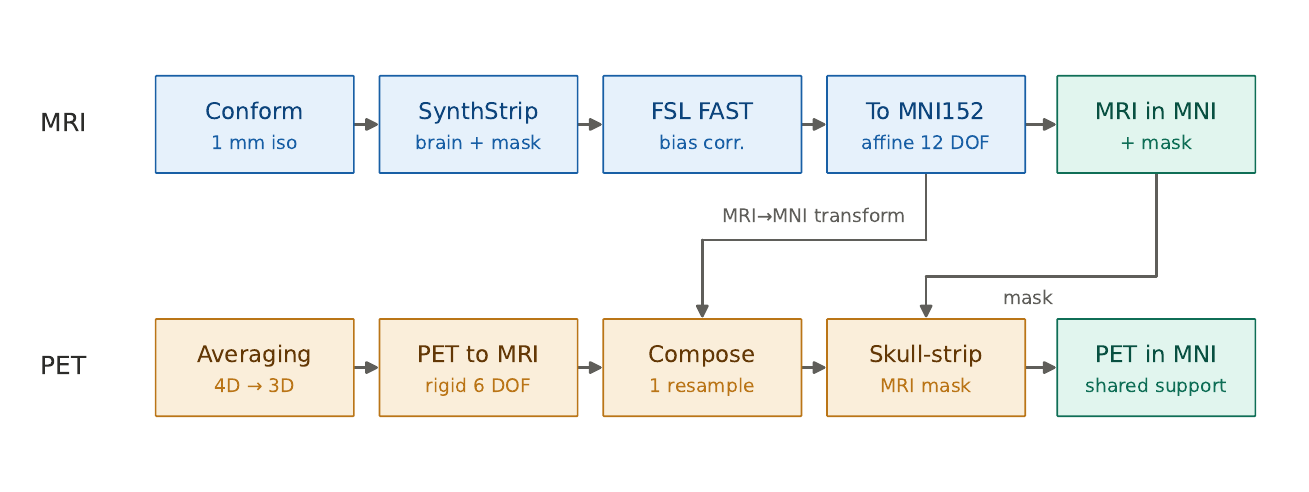}
    \caption{Overview of the preprocessing pipeline for MRI and PET.}
    \label{fig:pipeline}
\end{figure}

\begin{figure}[H]
    \centering
    \includegraphics[width=\textwidth]{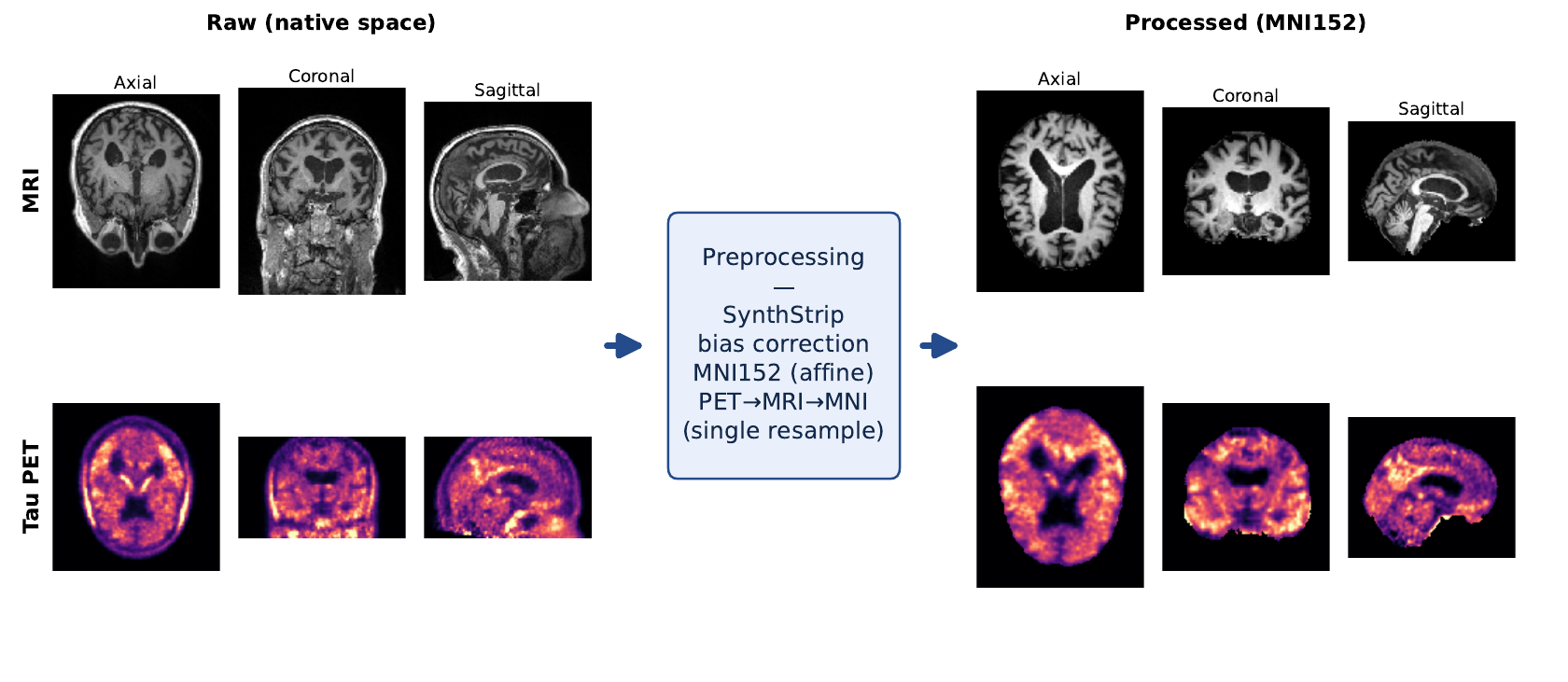}
    \caption{Effect of the preprocessing pipeline on a representative AD
    subject, for structural MRI (top) and Tau PET (bottom). For each modality,
    three orthogonal views (axial, coronal, sagittal) are shown before
    (left, native acquisition space) and after (right, MNI152 space)
    preprocessing.}
    \label{fig:preprocessing}
\end{figure}

Using the MRI as the registration target is more convenient since the structural image provides a far more stable and detailed anatomical reference than PET. This improves the accuracy and reproducibility of the spatial normalisation of the functional data. Figure~\ref{fig:pipeline} summarizes the full pipeline for both modalities while Figure~\ref{fig:preprocessing} shows the effect of the full pipeline on a representative subject.

\subsection{ROI feature extraction}

From spatially normalized volumes of the brain, the features of the region-of-interest are extracted using the Harvard-Oxford cortical and subcortical atlas distributed with FSL~\cite{makris2006harvardoxford, desikan2006parcellation, frazier2005limbic, goldstein2007hypothalamic}, which provides a standard reproducible anatomical parcellation of the brain. Features are computed directly from the preprocessed images in MNI space ensuring that the ROI representation is consistent with the imaging pipeline described above.

For each region, robust intensity summaries are computed, together with quality-control indicators describing the coverage and reliability of each region in the individual image. The complete set of per-region features (intensity statistics, robust $z$-scores, global-normalised ratios and quality-control indicators) is listed in Appendix~\ref{app:roi-features}.

\subsection{Plasma and tabular feature preparation}

In addition to the imaging modalities, plasma biomarkers and demographic variables are assembled into a subject-level tabular representation. The available plasma markers include phosphorylated-tau and amyloid-ratio measurements, complemented by age and sex. Complete plasma information is available for $712/881$ subjects ($80.8\%$), so an explicit availability indicator is retained and missing values are handled within the modelling pipeline rather than by ad hoc imputation of the raw table.

Crucially, all data-dependent preprocessing operations (imputation of missing values, feature scaling and categorical encoding) are encapsulated within a single modelling pipeline and fitted exclusively on the training portion of each fold. No statistic used for scaling or imputation is ever estimated on test subjects, which prevents information from the evaluation set leaking into the feature preparation~\cite{cawley2010overfitting}. The resulting subject-level feature tables, combining ROI summaries, plasma biomarkers and demographic variables, constitute the input to the tabular ROI/plasma branch and, through their out-of-fold predictions, to the multimodal fusion described in Section~\ref{sec:methods}.
\section{Models and Experimental Methodology}
\label{sec:methods}

\subsection{Problem formulation}

The task addressed in this work is a three-class, subject-level classification problem. Each subject is assigned to one of the diagnostic classes $\mathcal{Y} = \{\text{CN}, \text{MCI}, \text{AD}\}$, encoded as $y \in \{0, 1, 2\}$.

Let $s$ index subjects in the aligned $881$-subject cohort. Each subject is associated with up to four modality-specific representations: a Tau PET volume, a structural MRI volume, a tabular ROI-and-plasma feature vector and the inputs to a hierarchical PET-plasma model. A given branch $b$ produces, for subject $s$, a class-score vector
\begin{equation}
    \mathbf{p}^{(b)}_s = \big(p^{(b)}_{s,\text{CN}},\, p^{(b)}_{s,\text{MCI}},\, p^{(b)}_{s,\text{AD}}\big), \qquad \sum_{c} p^{(b)}_{s,c} = 1,
\end{equation}
and the final multimodal model combines the four branch vectors $\{\mathbf{p}^{(b)}_s\}_b$ into a single fused probability vector $\bar{\mathbf{p}}_s$, from which the subject-level decision is taken as $\hat{y}_s = \arg\max_c \bar{p}_{s,c}$. All models are trained and evaluated under the synchronized subject-level folds introduced in Section~\ref{sec:dataset}.

\subsection{Subject-level aggregation}

Because some subjects contribute more than one scan, the image-based branches first produce scan-level probabilities, which are then aggregated to a single probability vector per subject by averaging across that subject's scans:
\begin{equation}
    \mathbf{p}^{(b)}_s = \frac{1}{|\mathcal{S}_s|} \sum_{i \in \mathcal{S}_s} \mathbf{p}^{(b)}_{s,i},
\end{equation}
where $\mathcal{S}_s$ is the set of scans of subject $s$ and $\mathbf{p}^{(b)}_{s,i}$ is the softmax output for scan $i$. The argmax decision rule is then applied to the aggregated probabilities.

Subject-level aggregation ensures that the unit of prediction matches the unit of clinical interest (the subject, not the scan), it removes the dependence of the reported metrics on how many scans a subject happens to contribute and it is consistent with the subject-level fold definition. Since all scans of a subject share a single fold, the aggregated probability for a test subject is always produced by a model that never saw any of that subject's scans during training. These aggregated, out-of-fold probabilities are the quantities passed to the fusion stage.

\subsection{PET-only model}

The PET-only branch is a 3D image classifier operating directly on the preprocessed Tau PET volumes in MNI space. The backbone is a 3D DenseNet-121 model implemented within the MONAI-based medical imaging framework~\cite{huang2017densenet, cardoso2022monai}, producing three-class logits that are converted to class probabilities by a softmax function.

To address class imbalance, the network is trained with a class-balanced focal loss (CB-Focal), which combines focal loss~\cite{lin2018focalloss} with class-balanced reweighting based on the effective number of samples~\cite{cui2019classbalanced}. For subject scan \(i\), let \(\hat{\mathbf{p}}_i\) denote the softmax output and let \(\ell_i^{\mathrm{LS}}\) be the label-smoothed cross-entropy loss. The training loss is

\begin{equation}
    \mathcal{L}_{\text{CB-Focal}}
    =
    \frac{1}{N}
    \sum_{i=1}^{N}
    w_{y_i}\alpha_{y_i}
    \left(1-\exp(-\ell_i^{\mathrm{LS}})\right)^{\gamma}
    \ell_i^{\mathrm{LS}},
\end{equation}

where \(\gamma\) is the focusing parameter, \(w_{y_i}\) is the class-balanced weight associated with the true class and \(\alpha_{y_i}\) is an additional manually specified per-class factor. The class-balanced weights are derived from the effective number of samples as

\begin{equation}
    \tilde{w}_k =
    \frac{1-\beta}{1-\beta^{n_k}},
    \qquad
    w_k =
    \frac{\tilde{w}_k}
    {\frac{1}{K}\sum_{j=1}^{K}\tilde{w}_j},
\end{equation}

with \(n_k\) the number of training samples in class \(k\), \(K=3\) the number of classes and \(\beta \in [0,1)\). In the final configuration, \(\beta=0.999\), \(\gamma=1.5\), label smoothing is set to \(0.01\) and the additional class factor is \(\alpha=[1.0,1.2,1.0]\), providing a mild extra emphasis on the MCI class.  Full details of the
training objective and the checkpoint-selection score are given in
Appendix~\ref{app:pet_only_loss}.

Training is performed on the available scan-level PET volumes from the training subjects. During validation and testing, scan-level probabilities are aggregated at subject level by averaging the class probabilities across all scans of the same subject. The network is optimized with AdamW~\cite{loshchilov2019adamw}, using gradient accumulation and gradient clipping. The best checkpoint is selected exclusively on the inner validation subjects using a cost-sensitive subject-level score that combines MCI F1, macro-F1 and balanced accuracy while penalizing severe CN--AD errors. The final subject-level decision is obtained by applying the argmax rule to the aggregated probabilities.

The PET-only model serves as an image-based baseline that captures the spatial Tau signal. As reported in Section~\ref{sec:results}, it provides a useful but limited contribution on its own, with comparatively weak MCI recall and is most valuable as one branch within the multimodal fusion.

\subsection{MRI-only model}

The MRI-only branch is a 3D image classifier operating on the preprocessed structural MRI volumes in MNI space. The model uses the same backbone used in the PET-only model (3D DenseNet-121 backbone implemented in MONAI), with three output logits corresponding to CN, MCI and AD. The logits are converted to class probabilities by a softmax function.

Unlike the PET branch, the MRI-only model is trained with a class-weighted cross-entropy loss:
\begin{equation}
    \mathcal{L}_{\text{CE}}
    =
    -\frac{1}{N}
    \sum_{i=1}^{N}
    w_{y_i}\log \hat{p}_{i,y_i},
    \qquad
    w_k
    =
    \frac{n_{\text{tot}}}{K\,n_k},
\end{equation}
where \(n_k\) is the subject-level count of class \(k\), \(K=3\) is the number of classes and \(n_{\text{tot}}\) is the total number of subjects used to define the class weights. During training, each epoch is constructed in a subject-balanced way by randomly selecting one scan per training subject and shuffling the resulting scan set. This prevents subjects with multiple scans from contributing disproportionately to the optimization.

Model selection is performed using the subject-level predictions on the validation fold. Scan-level probabilities are first averaged across all scans of the same subject and the validation score is computed as
\begin{equation}
    S_{\text{MRI}}
    =
    \frac{1}{2}\,\mathrm{BalAcc}
    +
    \frac{1}{2}\,F1_{\text{macro}}.
\end{equation}
The primary selection mode is based on the argmax predictions obtained from the aggregated probabilities. Additional CN and AD probability thresholds are tuned and saved as diagnostics, but they are not used for the out-of-fold predictions passed to the stacking procedure.

The MRI branch plays a dual role: As a standalone model, it captures structural neurodegeneration and contributes strong CN recognition, but remains relatively CN-oriented and less balanced across classes, with weak MCI recognition; As a fusion branch, it nonetheless provides information complementary to the molecular Tau PET signal and to the systemic information carried by plasma biomarkers and its out-of-fold probabilities are one of the inputs to the multimodal fusion.

\subsection{ROI and plasma tabular model}
\label{sec:methods-tabular}

The tabular branch operates on a compact subject-level representation combining Harvard-Oxford ROI summaries, plasma biomarkers, demographic variables (age and sex) and optional quality-control indicators. The input table is first aggregated at subject level, so that each subject contributes a single feature vector and the outer evaluation remains aligned with the synchronized subject-level folds.

Within each outer training fold, ROI features are ranked by a univariate discriminative score computed only on the training subjects and the top-\(k\) ROI subset is retained before model fitting. Concretely, the score is the one-way ANOVA \(F\)-statistic, i.e. the ratio between the between-class and the
within-class variance of each feature:
\begin{equation}
    F(x)
    =
    \frac{\sum_{c} n_c \,(\bar{x}_c - \bar{x})^2}
         {\sum_{c} \sum_{i \in c} (x_i - \bar{x}_c)^2},
    \label{eq:anova_f}
\end{equation}
where \(\bar{x}_c\) and \(n_c\) are the mean and the size of class \(c\) and \(\bar{x}\) is the global mean; features with the largest \(F\) are the most discriminative across the three diagnostic groups. The feature configuration, model family and hyperparameters are then chosen by nested inner cross-validation (the inner selection score is given in Appendix~\ref{app:tabular_selection_score}). All data-dependent preprocessing steps, including imputation, scaling and one-hot encoding, are encapsulated inside a scikit-learn pipeline fitted exclusively on the training portion of each split~\cite{pedregosa2011sklearn}. Within this pipeline, missing plasma and ROI entries are median-imputed and categorical variables are most-frequent-imputed, with both statistics estimated on the training fold only, moreover, the plasma-availability flag is kept as an additional input feature.
When ROI\(\times\)plasma interaction terms are used, they are constructed only from ROI features selected within the corresponding training fold, preventing information from validation or test subjects from entering feature construction.

The candidate model families are an elastic-net regularized multinomial logistic regression and an ExtraTrees classifier~\cite{geurts2006extratrees}. The logistic-regression candidates are evaluated over the full set of predefined feature ablations, regularization strengths and \(\ell_1\)-ratios, whereas the ExtraTrees candidates are restricted to a smaller set of compact ROI-based configurations to limit model complexity and computational cost. For each outer fold, the three best inner-CV configurations are refitted on the outer-training subjects and combined into a weighted probability ensemble; the ensemble weights are proportional to the inner-validation scores of the selected configurations and normalized to sum to one, yielding a convex combination of the member probabilities. The resulting ensemble produces the out-of-fold subject-level probabilities used both for standalone evaluation and as one of the four inputs to the multimodal fusion.

The motivation for a tabular branch is twofold. First, ROI and plasma features provide a compact representation relative to voxel-level 3D images, which is well suited to the moderate-sample regime of multimodal neuroimaging studies and reduces the risk of overfitting compared with high-capacity image models.
Second, the features are interpretable in terms of anatomical regions, measurable plasma biomarkers and basic demographic covariates. As reported in Section~\ref{sec:results}, this branch is the strongest individual branch in terms of both overall accuracy and macro-averaged AUC, while remaining computationally light and not requiring end-to-end 3D image processing.

\subsection{Hierarchical PET-plasma model}

Before the subject-level fusion of all four branches (Section~\ref{subsec:fusion}), the hierarchical branch performs an earlier fusion of PET and plasma within a two-stage decision structure, which is one of the methodological contributions of this work. It reflects the observation that CN/MCI/AD classification is not a uniform multiclass problem, but involves two structurally different decision boundaries. It therefore decomposes the task into two binary stages, each combining a MONAI 3D DenseNet-121 PET model with a plasma-based logistic meta-model. \\

\begin{itemize}
    \item \textbf{Stage 1 (CN vs.\ non-CN):} the PET image model is trained as a two-class classifier separating CN from non-CN subjects. The PET branch is optimized with a focal loss on the two-class logits, with class weights derived from the inverse class proportions and focusing parameter \(\gamma=1.5\). A plasma model is trained on the same binary target and a logistic meta-model combines PET and plasma-derived scores to produce the final stage probability \(\hat{p}_1 = P(\text{non-CN}\mid x)\).

    \item \textbf{Stage 2 (MCI vs.\ AD):} the PET image model is trained on true MCI and AD subjects only, using a combined binary loss
    \[
        \mathcal{L}_{\text{Stage 2}}
        =
        0.75\,\mathcal{L}_{\text{BCE}}
        +
        0.25\,\mathcal{L}_{\text{Focal}},
    \]
    where the positive class is AD, the binary focal term uses focusing parameter \(\gamma=1.0\) (against \(\gamma=1.5\) in Stage~1) and the positive-class weight is proportional to \(n_{\text{MCI}}/n_{\text{AD}}\). As in Stage 1, PET and plasma scores are combined through a logistic meta-model, producing \(\hat{p}_2 = P(\text{AD}\mid x)\).\\
\end{itemize}

The three-class probability vector is obtained by hierarchical recomposition:
\begin{equation}
    \begin{aligned}
        P(\text{CN})  &= 1 - \hat{p}_1, \\
        P(\text{AD})  &= \hat{p}_1 \,\hat{p}_2, \\
        P(\text{MCI}) &= \hat{p}_1 \,(1 - \hat{p}_2).
    \end{aligned}
\end{equation}

Although Stage 2 is trained and selected only on MCI and AD subjects, it is inferred for all validation and test subjects, including CN subjects, so that the second-stage scores do not implicitly encode the true class. The final class probabilities are therefore available for every subject. The final class decision is then obtained through validation-tuned stage thresholds: subjects with Stage 1 probability below the CN/non-CN threshold are assigned to CN; among the remaining subjects, AD is assigned only when the Stage 2 AD probability exceeds its threshold and the Stage 1 non-CN probability also satisfies the AD gate. Otherwise, the subject is assigned to MCI.

The implementation is explicitly leakage-controlled. The outer cross-validation is subject-level, inner-validation subjects are drawn only from outer-training subjects, PET model selection uses only inner train and validation subjects, PET refitting uses only outer-training subjects, plasma and stage meta-models are fitted only on training subjects and the final stage thresholds are tuned only on inner-validation subjects. Outer test subjects are used exactly once, to produce the final out-of-fold prediction. Further details on the stage-specific losses, PET-plasma meta-models and threshold-selection procedure are provided in Appendix~\ref{app:hierarchical_details}. This branch is particularly useful for the MCI class, on which it achieves the best behaviour among the base models, although its thresholded two-stage decision rule also leads to higher fold-to-fold variability.

\subsection{Multimodal fusion}
\label{subsec:fusion}

The four branches, tabular ROI/plasma (TAB), structural MRI (MRI), PET-only (PETONLY) and hierarchical PET--plasma (HIER), are combined using only their synchronized out-of-fold predictions, so that the fusion never observes in-sample predictions from any branch. For each subject $s$, the inputs form a tensor of branch score vectors $\mathbf{P}_s \in \mathbb{R}^{4 \times 3}$ (branches $\times$ classes). The TAB, MRI and PETONLY branches contribute their probability outputs; the HIER branch contributes its thresholded decision, soft-encoded as a probability-like score vector, reflecting the discrete nature of its two-stage rule.

\paragraph{Primary model: fixed convex fusion.}
The primary multimodal model is a fixed, TAB-anchored convex combination of the four branch scores,
\begin{equation}
    \bar{\mathbf{p}}_s
    = 0.55\,\mathbf{p}^{(\text{TAB})}_s
    + 0.15\,\mathbf{p}^{(\text{MRI})}_s
    + 0.15\,\mathbf{p}^{(\text{PETONLY})}_s
    + 0.15\,\mathbf{p}^{(\text{HIER})}_s,
    \label{eq:fixed_fusion}
\end{equation}

with the final decision $\hat{y}_s = \arg\max_{c} \bar{p}_{s,c}$. The weights in Equation~\eqref{eq:fixed_fusion} are non-negative and sum to one, so the result is a valid convex combination. Crucially, these weights are fixed design choices, decided a priori on the basis of the relative strength of the branches (the tabular branch being the strongest single modality in terms of accuracy and
macro-averaged AUC) and are not fitted on the evaluation data. By fixing the weights, the fusion introduces no additional parameters estimated from the folds, which removes any risk of the fusion overfitting the evaluation set and gives a simple, transparent and reproducible decision rule.

\begin{figure}[H]
    \centering
    \includegraphics[width=0.85\textwidth]{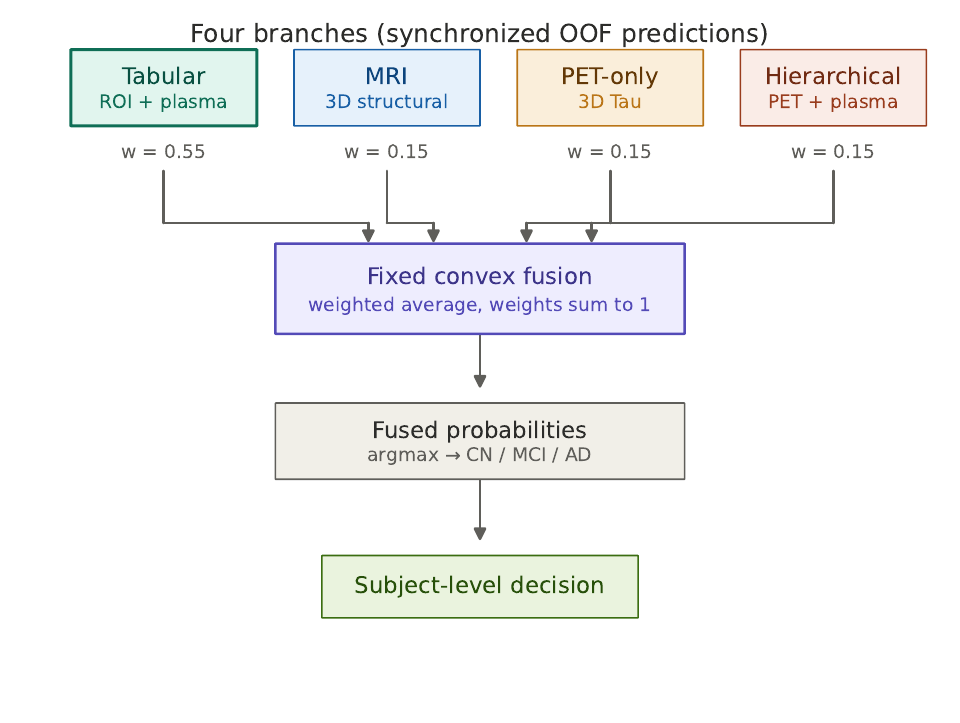}
    \caption{Multimodal late-fusion architecture. The fixed convex fusion combines them through a weighted average with a priori
    weights $[0.55, 0.15, 0.15, 0.15]$ and the subject-level decision is taken as the argmax of the fused probabilities.}
    \label{fig:fusion}
\end{figure}

\paragraph{Robustness analysis: nested convex superlearner.}
To verify that the fixed weights are a sound choice rather than an arbitrary one, a learned alternative is examined as a robustness analysis rather than as the final model. Following the super learner principle~\cite{vanderlaan2007superlearner, wolpert1990stackedgeneralization}, the convex weights are instead estimated from the out-of-fold predictions. To enforce non-negativity and the sum-to-one constraint, the weights are parametrized through a softmax, $\mathbf{w} = \mathrm{softmax}(\mathbf{z})$ and estimated by minimizing a class-balanced negative log-likelihood in which each sample is reweighted by the inverse frequency of its class (normalised to unit mean), so that the minority classes are not dominated by CN. A quadratic regularization term anchors the learned weights towards a reference anchor vector $\mathbf{a}$:
\begin{equation}
    \min_{\mathbf{z}} \;
    \mathcal{L}_{\text{CB-NLL}}\big(y, \bar{\mathbf{p}}(\mathbf{z})\big)
    \;+\; \lambda \, \lVert \mathbf{w}(\mathbf{z}) - \mathbf{a} \rVert^2,
    \label{eq:nested_obj}
\end{equation}

where $\lambda \ge 0$ controls the anchor strength. Two fusion families are considered: a global variant, with a single weight vector $\mathbf{w} \in \mathbb{R}^4$ shared across classes and a classwise variant, with a separate convex weight vector per class $\mathbf{W} \in \mathbb{R}^{3 \times 4}$, so that each class is allowed its own mixture of branches.

The fusion family and the regularization strength $\lambda$ (selected from $\{0, 0.01, 0.03, 0.1, 0.3, 1, 3, 10\}$) are chosen by nested cross-validation. Within each outer training fold, an inner cross-validation over the synchronized folds selects the configuration that maximizes a class-balanced selection score combining accuracy, macro-F1, balanced accuracy, MCI and AD F1, macro AUC and average precision, while penalizing severe CN$\leftrightarrow$AD confusions; the selected configuration is refitted on the outer-training folds and applied once to the held-out outer fold. This nested protocol avoids selecting hyperparameters on the same data used for evaluation, a well-documented source of optimistic bias~\cite{cawley2010overfitting}. Full details of the objective and the weight estimation are given in Appendix~\ref{app:superlearner_details}. The sensitivity of the result to the anchor is assessed by repeating the whole procedure over the three anchors above, from the uniform anchor to the TAB-anchored one. As reported in Section~\ref{sec:results}, the learned weights remain close to the fixed configuration of Equation~\eqref{eq:fixed_fusion} and do not yield a meaningful improvement over it, which supports the use of the fixed convex fusion as the primary model.

\paragraph{Logistic meta-learner.}
As a further, more expressive fusion baseline, a multinomial logistic meta-learner is trained on a set of diagnostics derived from the branch probability vectors (per-branch class probabilities, confidence, margin and entropy, inter-branch agreement and class-wise voting statistics). Its inverse regularization strength $C$ (from $\{0.03, 0.1, 0.3, 1, 3, 10\}$) and class weights are selected by the same nested inner cross-validation, again using only out-of-fold predictions, and no posterior probability calibration is applied to its outputs; the effect of such calibration is examined separately as an ablation (Appendix~\ref{app:calibration_ablation}). It is reported not as a competing primary model, but as a more expressive, MCI-oriented alternative to the fixed convex fusion, whose results are discussed in Section~\ref{sec:results}.

\subsection{Evaluation metrics}

All models are evaluated on the aggregated, out-of-fold, subject-level predictions over the aligned cohort. Given the class imbalance, no single metric is sufficient and a panel of complementary metrics is reported.

Overall accuracy measures the fraction of correctly classified subjects but is dominated by the majority CN class. Balanced accuracy, the mean of the per-class recalls and macro-F1, the unweighted mean of the per-class F1 scores, give equal weight to each class and are therefore more informative under imbalance; weighted-F1 is also reported. To characterise behaviour on individual classes (in particular on MCI) class-wise precision, recall and F1 are reported, with special attention to the MCI F1 score ($\text{F1}_{\text{MCI}}$). The discriminative ability of the probabilistic outputs is summarised by the multiclass area under the ROC curve (AUC), computed in a one-vs-rest, macro-averaged form.

Finally, confusion matrices are used to analyse the structure of the errors, distinguishing clinically mild confusions (e.g.\ MCI$\rightarrow$CN) from severe ones (CN$\leftrightarrow$AD) and to support the subsequent error analysis. Pairwise differences between models are assessed with exact McNemar tests on the subject-level decisions and $95\%$ confidence intervals for the reported metrics are obtained by subject-level bootstrap resampling with $2000$ replicates. The effect of post-hoc probability calibration of the logistic meta-learner is examined as an ablation, reported in Appendix~\ref{app:calibration_ablation}.
\section{Results}
\label{sec:results}

All branches and the fusion are evaluated on the aggregated, out-of-fold, subject-level predictions over the synchronized $881$-subject cohort. For each model, the main text reports the pooled confusion matrix together with the most relevant headline metrics; the complete panels of pooled and fold-level metrics (accuracy, balanced accuracy, macro- and weighted-F1, per-class F1 and recall, macro AUC and AP, log-loss and ECE) are collected in Appendix~\ref{app:appendixB}.

\subsection{PET-only branch}
\label{subsec:results-petonly}

The PET-only branch reaches a pooled accuracy of $\mathbf{0.623}$, balanced accuracy $\mathbf{0.606}$ and macro-F1 $\mathbf{0.580}$, with a macro one-vs-rest AUC of $\mathbf{0.746}$. Its per-class F1 is $0.742$ (CN), $0.411$ (MCI) and $0.588$ (AD): the molecular Tau signal has the second-best MCI recall among the single branches ($\mathbf{0.397}$) and a solid AD recall ($0.691$). The pooled confusion matrix is given in Table~\ref{tab:pet-cm} and the full metric panel in Appendix~\ref{app:full_metrics_pet}, Table~\ref{tab:pet-overall-app}.

\begin{table}[H]
    \centering
    \caption{Pooled subject-level confusion matrix of the PET-only branch. Rows correspond to true classes and columns to predicted classes.}
    \label{tab:pet-cm}
    \begin{tabular}{lccc}
    \toprule
     & $\widehat{\mathrm{CN}}$ & $\widehat{\mathrm{MCI}}$ & $\widehat{\mathrm{AD}}$ \\
    \midrule
    \textbf{CN}  & $374$ & $120$ & $18$ \\
    \textbf{MCI} & $118$ & $108$ & $46$ \\
    \textbf{AD}  & $4$   & $26$  & $67$ \\
    \bottomrule
    \end{tabular}
\end{table}

\subsection{MRI-only branch}
\label{subsec:results-MRI}

The structural MRI branch reaches a pooled accuracy of $\mathbf{0.649}$, balanced accuracy $\mathbf{0.577}$ and macro-F1 $\mathbf{0.562}$, with a macro AUC of $\mathbf{0.718}$. It behaves as a CN-oriented detector, attaining the highest CN recall of all branches ($0.857$) but the lowest MCI recall ($\mathbf{0.276}$) and its per-class F1 is $0.773$ (CN), $0.361$ (MCI) and $0.552$ (AD). The pooled confusion matrix is given in Table~\ref{tab:mri-cm}, with the full metric panel in Appendix~\ref{app:full_metrics_mri},
Table~\ref{tab:mri-overall-app}.

\begin{table}[H]
    \centering
    \caption{Pooled subject-level confusion matrix of the structural MRI branch. Rows correspond to true classes and columns to predicted classes.}
    \label{tab:mri-cm}
    \begin{tabular}{lccc}
    \toprule
     & $\widehat{\mathrm{CN}}$ & $\widehat{\mathrm{MCI}}$ & $\widehat{\mathrm{AD}}$ \\
    \midrule
    \textbf{CN}  & $439$ & $52$ & $21$ \\
    \textbf{MCI} & $163$ & $75$ & $34$ \\
    \textbf{AD}  & $22$  & $17$ & $58$ \\
    \bottomrule
    \end{tabular}
\end{table}

\subsection{Tabular ROI-plasma branch}
\label{sec:results-tabular}

The tabular ROI-plasma branch is the strongest single model overall. It reaches a pooled accuracy of $\mathbf{0.671}$ and the highest macro one-vs-rest AUC of all branches ($\mathbf{0.791}$), together with the lowest fold-to-fold accuracy variability ($\pm 0.016$). Its per-class F1 is $0.806$ (CN), $0.400$ (MCI) and $0.563$ (AD), and the per-class OvR AUC is $0.801$ (CN), $0.648$ (MCI) and $0.922$ (AD). It shares with the image-derived branches the dominant MCI$\rightarrow$CN error mode. The pooled confusion matrix is given in Table~\ref{tab:tab-cm}, with the full metric panel in Appendix~\ref{app:full_metrics_tab}, Table~\ref{tab:tab-overall-app}.

\begin{table}[H]
    \centering
    \caption{Pooled subject-level confusion matrix of the tabular ROI--plasma branch. Rows correspond to true classes and columns to predicted classes.}
    \label{tab:tab-cm}
    \begin{tabular}{lccc}
    \toprule
     & $\widehat{\mathrm{CN}}$ & $\widehat{\mathrm{MCI}}$ & $\widehat{\mathrm{AD}}$ \\
    \midrule
    \textbf{CN}  & $439$ & $63$  & $10$ \\
    \textbf{MCI} & $134$ & $92$  & $46$ \\
    \textbf{AD}  & $4$   & $33$  & $60$ \\
    \bottomrule
    \end{tabular}
\end{table}

A post-hoc inspection of the univariate ROI ranking confirms that the discriminative tabular signal is anatomically coherent with the known topography of Alzheimer's disease (Appendix~\ref{app:full_metrics_tab}, Table~\ref{tab:roi-selection}). Across all five outer folds the selection consistently retained the medial-temporal structures (the medial-temporal-lobe macro-region and the left and right amygdala), the posterior default-mode regions (precuneus and posterior cingulate gyrus), the lateral temporal cortex (posterior and temporo-occipital middle temporal gyrus) and the inferior parietal cortex (angular gyrus), together with the tau- and amyloid-signature macro-regions; the single highest-ranked feature in every fold was the robust tau-AD-signature uptake (\texttt{macro\_tau\_ad\_signature\_like}, $p90$ over global median). The hippocampus proper was selected less systematically (left hippocampus in three folds, right in one), consistent with its signal being largely subsumed by the medial-temporal macro-region and with the amygdala carrying a strong share of the medial-temporal tau signal in AV1451 imaging. This recovery of the canonical AD signature is a sanity check on the tabular branch, the model's strength rests on biologically plausible regional features.

\subsection{Hierarchical PET--plasma branch}
\label{sec:results-hier}

The hierarchical PET-plasma branch is evaluated using final threshold-based subject-level decisions, while probabilistic metrics use the recomposed class probabilities. It attains the best MCI behaviour of all single branches (F1$_{\mathrm{MCI}}=\mathbf{0.455}$, MCI recall $\mathbf{0.474}$) and the best calibration (ECE $=0.040$), at the cost of the highest fold-to-fold variability among all branches (accuracy std 0.092; full per-metric standard deviations in Appendix~\ref{app:full_metrics_hier}, Table~\ref{tab:hier-overall-app}), a direct consequence of its threshold-based two-stage rule. Its pooled accuracy is $\mathbf{0.628}$, balanced accuracy $\mathbf{0.610}$ and macro AUC $\mathbf{0.760}$. The pooled confusion matrix is given in Table~\ref{tab:hier-cm}, with the full metric panel in Appendix~\ref{app:full_metrics_hier}, Table~\ref{tab:hier-overall-app} and the stage-wise binary
performance in Appendix~\ref{app:full_metrics_hier}, Table~\ref{tab:hier-stages-app}.

\begin{table}[H]
\centering
\caption{Pooled subject-level confusion matrix of the hierarchical PET--plasma branch. Rows correspond to true classes and columns to predicted classes.}
\label{tab:hier-cm}
\begin{tabular}{lccc}
    \toprule
    & $\widehat{\mathrm{CN}}$ & $\widehat{\mathrm{MCI}}$ & $\widehat{\mathrm{AD}}$ \\
    \midrule
    \textbf{CN}  & $361$ & $135$ & $16$ \\
    \textbf{MCI} & $106$ & $129$ & $37$ \\
    \textbf{AD}  & $3$   & $31$  & $63$ \\
    \bottomrule
\end{tabular}
\end{table}

\subsection{Multimodal fusion}
\label{sec:results-fusion}

The primary multimodal model is the fixed convex fusion, a TAB-anchored convex combination of the four branch score vectors with weights $\mathbf{w}=[0.55,0.15,0.15,0.15]$, operating exclusively on the synchronized out-of-fold branch predictions. It reaches a pooled accuracy of $\mathbf{0.703}$, balanced accuracy $\mathbf{0.660}$ and macro-F1 $\mathbf{0.652}$, with per-class F1 of $0.809$ (CN), $\mathbf{0.496}$ (MCI) and $0.646$ (AD) and a macro AUC of $\mathbf{0.804}$. Its dominant residual error is the MCI$\rightarrow$CN confusion ($116$ subjects), consistent with the difficulty of the MCI class across all branches. The pooled confusion matrix is given in Table~\ref{tab:fusion-cm-fixed}, with the full metric panel in Appendix~\ref{app:full_metrics_fusion}, Table~\ref{tab:fusion-overall-app}.

\begin{table}[H]
    \centering
    \caption{Pooled subject-level confusion matrix of the fixed convex fusion. Rows correspond to true classes and columns to predicted classes.}
    \label{tab:fusion-cm-fixed}
    \begin{tabular}{lccc}
    \toprule
     & $\widehat{\mathrm{CN}}$ & $\widehat{\mathrm{MCI}}$ & $\widehat{\mathrm{AD}}$ \\
    \midrule
    \textbf{CN}  & $429$ & $74$  & $9$  \\
    \textbf{MCI} & $116$ & $123$ & $33$ \\
    \textbf{AD}  & $3$   & $27$  & $67$ \\
    \bottomrule
    \end{tabular}
\end{table}

\begin{figure}[H]
    \centering
    \includegraphics[width=\textwidth]{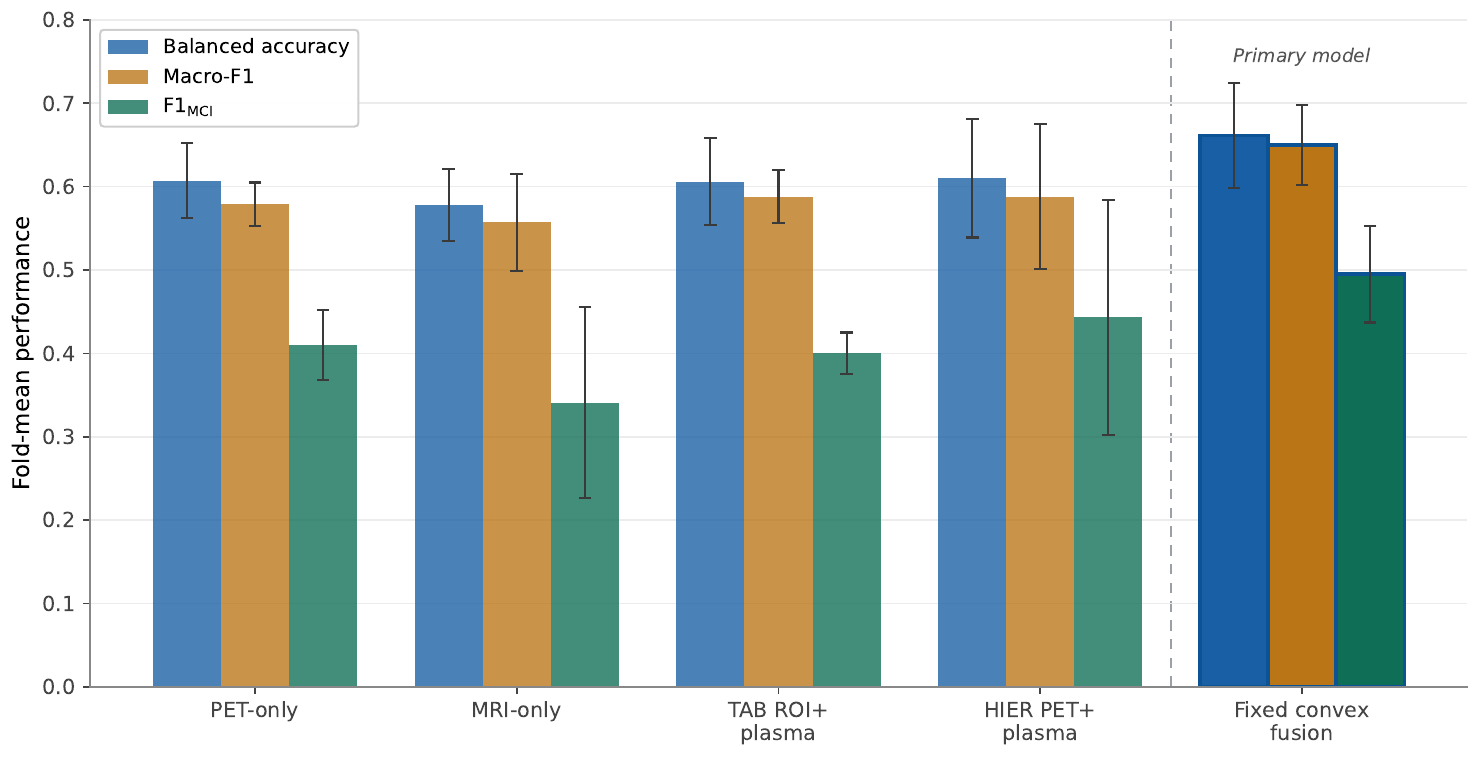}
    \caption{Subject-level comparison of the four branches and the primary fixed convex fusion on the synchronized $881$-subject cohort. Bars report the fold-mean balanced accuracy, macro-F1 and MCI F1; error bars denote the standard deviation across the five outer folds.}
    \label{fig:results-comparison}
\end{figure}

An exact McNemar test on the subject-level decisions confirms that the fusion differs significantly from every single branch (against the tabular branch $p=0.04$; against MRI $p=0.009$; against PET-only and the hierarchical branch $p<10^{-7}$) and the corresponding $95\%$ bootstrap confidence intervals for the pooled metrics (accuracy $[0.672, 0.732]$, macro-F1 $[0.616, 0.688]$) exclude the
single-branch values. The improvement is therefore not attributable to noise. Figure~\ref{fig:results-comparison} summarizes the comparison across all four branches and the fusion on the three metrics most relevant under class imbalance. A side-by-side view of all five pooled confusion matrices is provided in Appendix~\ref{app:cm_grid}, Figure~\ref{fig:cm-grid}.

\paragraph{Robustness of the fixed weights.}
To verify that the fixed weights are a sound choice rather than an arbitrary one, the convex weights were also learned from the out-of-fold predictions by the nested superlearner of Section~\ref{subsec:fusion}, under the same tabular anchor. The learned weights converge to the fixed configuration. In the two folds where the global family is selected, the estimated weights are $[0.550, 0.151, 0.150, 0.149]$ and $[0.551, 0.148, 0.152, 0.149]$, essentially indistinguishable from the fixed $[0.55, 0.15, 0.15, 0.15]$; in the three folds where the classwise family is selected, the tabular branch remains the dominant contributor for every class (per-class $w_{\mathrm{TAB}}$ in the range $0.53$--$0.60$), with the remaining mass spread over the MRI, PET and hierarchical branches. The resulting performance is statistically indistinguishable from the fixed fusion (pooled accuracy $0.697$ versus $0.703$, macro-F1 $0.647$ versus $0.652$, MCI F1 $0.489$ versus $0.496$), the two pooled confusion matrices differing by at most three subjects per cell. The data-driven optimum therefore reproduces the a-priori weights rather than improving on them, which confirms the fixed, parameter-free convex fusion as the more parsimonious choice and the primary model of this work.

\paragraph{MCI-oriented operating point.}
The fixed fusion balances overall accuracy and MCI sensitivity. Because mis-classifying an MCI subject as cognitively normal is clinically more consequential than the reverse, an alternative operating point that explicitly favours MCI sensitivity is also reported, obtained with the logistic meta-learner trained with strong MCI class weights selected by nested inner cross-validation. As summarised in Table~\ref{tab:fusion-mci-oriented}, this variant raises MCI recall from $0.452$ to $0.537$ and MCI F1 to $0.507$, the highest among all fusion models, while attaining the highest macro AUC ($0.810$) and the fewest severe CN$\leftrightarrow$AD confusions ($5$ against $12$). The gain in MCI sensitivity is obtained at the expense of overall accuracy, which decreases to $0.672$ and of CN recall, as more CN subjects are assigned to MCI. The full metric panel and the pooled confusion matrix are provided in Appendix~\ref{app:full_metrics_logistic_meta}. The two configurations represent two points on the same accuracy-sensitivity trade-off, whose clinical interpretation is discussed in Section~\ref{sec:discussion}.

\begin{table}[H]
    \centering
    \caption{MCI-oriented operating point (logistic meta-learner with strong MCI class weights), compared with the primary fixed convex fusion. Pooled OOF metrics on the $881$-subject cohort.}
    \label{tab:fusion-mci-oriented}
    \begin{tabular}{lcc}
    \toprule
    \textbf{Metric} & \textbf{Fixed fusion} & \textbf{MCI-oriented meta} \\
    \midrule
    Accuracy            & $\mathbf{0.703}$ & $0.672$ \\
    Balanced accuracy   & $\mathbf{0.660}$ & $0.614$ \\
    Macro-F1            & $\mathbf{0.652}$ & $0.630$ \\
    \midrule
    F1$_{\mathrm{MCI}}$ & $0.496$ & $\mathbf{0.507}$ \\
    Recall$_{\mathrm{MCI}}$ & $0.452$ & $\mathbf{0.537}$ \\
    Recall$_{\mathrm{CN}}$  & $\mathbf{0.838}$ & $0.770$ \\
    \midrule
    AUC (macro OvR)     & $0.804$ & $\mathbf{0.810}$ \\
    Severe CN$\leftrightarrow$AD & $12$ & $\mathbf{5}$ \\
    \bottomrule
    \end{tabular}
\end{table}
\section{Discussion}
\label{sec:discussion}

This section interprets the results of Section~\ref{sec:results}. It begins with the behaviour of the individual branches and the design choices behind them, turns to the multimodal fusion that combines them and closes with a set of general considerations on the central difficulty of the MCI class, on the discrete formulation of the task and on the directions they suggest.

\paragraph{Single-branch results}

The four branches expose a consistent and clinically meaningful pattern. The CN and AD classes are comparatively separable, whereas MCI is the hardest class for every branch. Across branches the MCI one-vs-rest AUC stays well below the AD one and the dominant error mode is always the MCI$\rightarrow$CN confusion. This is the signature of MCI as an intermediate, heterogeneous condition. Many MCI subjects are biologically close to the cognitively normal end of the spectrum and the available signal does not place them unambiguously on one side of the CN/MCI boundary.\\

The branches differ in where they are reliable and this non-overlap is what later makes their fusion worthwhile. The tabular ROI-plasma branch is the strongest single model overall, with the highest accuracy ($0.671$), the highest macro-averaged AUC ($0.791$) and the lowest fold-to-fold variability ($\pm 0.016$ in accuracy). Its compact, interpretable ROI-and-plasma representation is well matched to the moderate-sample regime and creates a stable, well-ranked classifier. The structural MRI branch attains the highest CN recall ($85.7\%$) but the lowest MCI recall ($27.6\%$), together with the largest number of severe CN$\leftrightarrow$AD confusions ($43$ subjects). It behaves as a CN-oriented detector, reliable at recognising preserved anatomy but conservative in committing to the pathological classes. The PET-only branch achieves the highest AD recall ($69.1\%$) and a substantially better MCI recall ($39.7\%$), in line with the fact that Tau-PET captures a molecular signal that becomes spatially salient already in early stages. The hierarchical PET-plasma branch attains the best MCI behaviour of all single branches (F1$_{\mathrm{MCI}}=0.455$, MCI recall $47.4\%$) and the best calibration (ECE $=0.040$), at the price of the highest fold-to-fold variability, which is a direct consequence of its threshold-based two-stage decision rule.\\

The two image branches are trained with deliberately different losses and this asymmetry responds to the different failure modes of the two modalities. The MRI branch uses a class-weighted cross-entropy, with each class reweighted inversely to its frequency ($w_k = n_{\text{tot}}/(K\,n_k)$) and combined with a subject-balanced sampler.
Rescaling the minority classes without distorting the geometry of the loss is a mild form of imbalance correction. The choice reflects an empirical observation, namely that the structural signal is informative mainly for the extremes of the spectrum (preserved versus atrophied anatomy) and that a stronger minority-focused objective tended to destabilise the already CN-oriented branch without a reliable gain on MCI. The PET branch instead uses a class-balanced focal loss based on the effective number of samples~\cite{cui2019classbalanced}, with focusing parameter $\gamma=1.5$. The focal term down-weights easy, confidently classified examples and concentrates the gradient on the hard, near-boundary subjects, while the class-balanced term corrects imbalance through the effective number of samples rather than the raw frequency. This stronger objective is justified by the nature of the Tau-PET signal, where the information that discriminates the pathological classes is concentrated in a relatively small number of hard, ambiguous subjects. The asymmetry between the two losses therefore mirrors the asymmetry between the two modalities.\\

The hierarchical PET-plasma branch was designed to reflect the asymmetric structure of the task, first separating CN from non-CN and then estimating the AD component within the non-CN decision. Its value is concentrated on the intermediate class. Relative to PET-only it raises MCI recall from $0.397$ to $0.474$, MCI F1 from $0.411$ to $0.455$ and the stage-wise analysis shows that the learned PET$+$plasma meta-model improves on either source alone (Stage~1 balanced accuracy $0.719$ versus $0.689$ for PET and $0.663$ for plasma; Stage~2 balanced accuracy $0.812$ with AD recall $0.714$). The Stage~2 AUC of the joint model ($0.905$) essentially matches PET alone ($0.906$). The meta-model does not improve the global ranking of AD probability, but it produces a more balanced operating point after the learned combination and the threshold-based decision.
The cost of this decision-aware design is the highest fold-to-fold variability of all branches (accuracy std $0.092$),
driven by the sensitivity of a two-stage thresholded rule to the small AD class. The branch is therefore not a standalone replacement for the others, but an MCI-oriented component whose error profile differs from the flat classifiers, which is exactly what makes it useful in fusion.

\paragraph{Multimodal fusion}
\label{sec:disc-fusion}
The central result of this work is that combining the four branches yields a measurable and statistically significant improvement over every single branch and that the largest part of this improvement is realised on the most difficult class.\\

The fixed convex fusion reaches accuracy $0.703$ and macro-F1 $0.652$, against $0.671$ and $0.590$ for the strongest single branch and the gain is most pronounced on MCI, whose F1 rises from $0.400$ to $0.496$ and whose recall rises from $0.338$ to $0.452$. The fusion, drawing CN stability from the MRI and tabular branches and MCI/AD sensitivity from the PET-based branches, shifts a substantial number of MCI subjects away from the CN prediction while keeping severe CN$\leftrightarrow$AD confusions low ($12$ subjects). However it pays a small calibration cost (pooled ECE $0.094$, the highest among all models), since averaging well-calibrated probabilities contracts the fused scores towards the centre. This does not affect the reported metrics, as the decision is taken by argmax.\\

A central methodological question is whether the fixed, a-priori weights $[0.55,0.15,0.15,0.15]$ leave performance on the table relative to weights learned from the data. The nested convex superlearner answers this question empirically. When the weights are estimated from the out-of-fold predictions, they converge back to the fixed configuration. In the folds where a single global weight vector is selected, the learned weights are $[0.550,0.151,0.150,0.149]$ and $[0.551,0.148,0.152,0.149]$ and even in the classwise folds the tabular branch remains dominant for every class. The resulting performance is statistically indistinguishable from the fixed fusion (accuracy $0.697$ versus $0.703$; an exact McNemar test gives $p=0.18$ on only nine discordant subjects). The data-driven optimum therefore reproduces the a-priori weights rather than improving on them, which both validates the fixed choice and justifies preferring it as a simpler parameter-free model.\\

Because misclassification of an MCI subject as cognitively normal is clinically more costly than the reverse, the logistic meta-learner with strong MCI class weights provides a useful alternative operating point. It raises MCI recall to $0.537$ and MCI F1 to $0.507$, the best MCI behaviour of all models, with the highest macro AUC ($0.810$) and the fewest severe CN$\leftrightarrow$AD confusions ($5$), at the cost of roughly three points of overall accuracy ($0.672$) and a lower CN recall. The fixed fusion and the meta-learner are therefore two points on the same accuracy-sensitivity frontier and the appropriate choice depends on the relative clinical cost assigned to missed MCI cases. That the two configurations span this frontier on identical inputs, differing only in how the branch outputs are weighted, is itself an informative characterisation of what the fused signal can and cannot deliver.\\

A practically relevant finding is that the multimodal performance is close to the ceiling extractable from these branches by fusion alone. Learning the weights did not help, calibrating the meta-learner did not help (Appendix~\ref{app:calibration_ablation}) and enriching the meta-learner with additional disagreement features did not improve on the simple convex combination. The system sits on an accuracy-sensitivity frontier on which no configuration simultaneously achieves high accuracy and high MCI recall. This indicates that the residual MCI difficulty is not a deficiency of the fusion
mechanism but a limit of the information content of the branches themselves, so further improvements must come from new information, not from a more elaborate combination of the existing signals.\\

\paragraph{General considerations}
\label{sec:disc-general}

Across every experiment, the recovery of the MCI class is the limiting factor. Even the most MCI-sensitive configuration recovers only slightly more than half of the MCI subjects and the residual error is overwhelmingly the MCI$\rightarrow$CN confusion. This matches the clinical reality that early MCI and normal ageing form a near-continuum and that a substantial fraction of MCI subjects carry little or no detectable pathological signal in the available modalities at the time of acquisition. The bottleneck is therefore not purely a modelling artefact, it reflects a genuine limit of the cross-sectional information available for the intermediate class. That every branch and the fusion fails in the same direction (towards CN) reinforces this reading.\\

The task is posed as a three-way classification into the discrete labels CN, MCI and AD. This formulation is convenient and matches the diagnostic categories used in practice, but it is in tension with the underlying biology. Alzheimer's disease is a continuous, progressive process and the CN/MCI/AD labels are partly arbitrary discretisations of that continuum, with MCI occupying the transition region between the two extremes. Forcing a hard assignment into three classes imposes sharp decision boundaries on subjects who lie near the CN/MCI or MCI/AD frontiers and it penalises a model equally for a near-boundary CN/MCI confusion (clinically minor) and for a CN/AD confusion (clinically severe). The metrics used here only partially mitigate this through the explicit reporting of severe CN$\leftrightarrow$AD confusions; they cannot fully capture the ordinal, graded structure of the problem. Much of the error that the models are penalised for is the discretisation of a continuum into three classes.\\

The synchronized cohort comprises $881$ subjects with a marked class imbalance ($512$ CN, $272$ MCI, $97$ AD) and all results are obtained on a single cohort. The small AD group in particular contributes to the high fold-to-fold variability of the AD and MCI metrics and external validation on an independent cohort is required before any claim of generalisation can be made.

\paragraph{Future directions}
\label{sec:disc-future}

The single most consequential direction suggested by these results is to abandon the discrete three-class target in favour of a formulation that respects the continuous nature of disease progression. The remaining directions follow from the same diagnosis, that the MCI bottleneck is a limit of information rather than of modelling.\\

A radical but promising reformulation is to recast the problem as a regression onto a continuous disease-severity score rather than onto discrete labels. A continuous target, for instance a cognitive or biomarker-derived progression index or a data-driven latent severity axis, would place each subject on a graded scale, would dissolve the artificial sharpness of the CN/MCI boundary that drives most of the residual error and would create a clinically richer output. The model would return a position on the progression continuum together with an associated uncertainty, rather than a single hard label. The discrete CN/MCI/AD decision could still be recovered when required by thresholding the predicted score, with thresholds chosen to reflect the asymmetric clinical cost of different error types.\\

A less radical but more easily actionable step is to replace the flat three-class objective with an ordinal one that preserves the ordering CN $<$ MCI $<$ AD and penalises errors by their distance along this axis, so that a CN/AD confusion is treated as strictly more costly than a CN/MCI confusion. This retains the familiar label structure while removing the assumption that the three classes are exchangeable and unordered and it aligns the training objective with the clinical cost structure that the present metrics can only report after the fact.\\

Closely related is the idea of treating MCI explicitly as the transition region between CN and AD rather than as a third homogeneous category. One could model the CN-to-AD axis as a latent continuum and interpret MCI as the high-uncertainty band along that axis. This is both more faithful to the biology and better matched to the empirical behaviour observed here, where MCI is precisely the region of maximal model uncertainty and of maximal overlap with the two extremes.\\

The present analysis is cross-sectional and the MCI bottleneck is due in part to the limited pathological signal available at a single time point. A trajectory is far more informative about transition than a snapshot, so incorporating longitudinal information, for example the rate of hippocampal atrophy or of Tau accumulation between visits, could be a natural way to improve MCI discrimination and fits directly into a regression-on-severity formulation as a prediction of the rate of progression. In the same spirit, additional biomarkers orthogonal to the imaging and plasma sources used here, in particular cerebrospinal-fluid measurements, could supply the signal that the cross-sectional imaging modalities lack for the intermediate class.\\

Finally, the high fold-to-fold variability of the molecular and hierarchical branches suggests that stability-oriented interventions, such as multi-seed ensembling of the base branches before fusion, are a low-risk way to reduce variance and improve the robustness of the downstream combination, independently of the reformulations above.
\section{Conclusion}
\label{sec:conclusion}

This work developed a rigorous, leakage-controlled framework for subject-level CN/MCI/AD classification by combining four complementary branches, a 3D Tau PET model, a 3D structural MRI model, a tabular ROI-and-plasma model and a hierarchical PET-plasma model, over a synchronized cohort of 881 ADNI subjects. The contribution lies not in the backbone architectures, which follow established designs, but in the hierarchical formulation of the classification task and in a systematic, leakage-controlled comparison and fusion of the four modalities that quantifies the diagnostic information each one carries. All branches were aligned at the subject level and evaluated under identical, no-leak cross-validation folds and the multimodal fusion was built exclusively on synchronized out-of-fold predictions.

Within this framework, the multimodal fusion improved significantly on every individual branch, with the largest gain realised on the most difficult class, MCI. A fixed, parameter-free convex combination served as the primary model and a nested superlearner confirmed its weights rather than improving on them, which validates the simpler choice. A logistic meta-learner provided a complementary, MCI-oriented operating point on the same accuracy--sensitivity frontier, making explicit the trade-off between overall accuracy and the recovery of the intermediate class.

The central and persistent finding is that MCI remains the bottleneck. Every branch and the fusion itself fails predominantly in the same direction, towards the cognitively normal class and no fusion configuration simultaneously achieves high accuracy and high MCI sensitivity. This points to a limit of the information available cross-sectionally for the intermediate class rather than to a deficiency of the fusion mechanism. The main limitations of the study follow from this and from its design. The results rest on a single, class-imbalanced cohort, so external validation on an independent dataset remains necessary before any claim of generalisation can be made and the small AD group in particular drives the fold-to-fold variability of the molecular and hierarchical branches.

These limitations also show how the most promising way forward is not a more elaborate combination of the same signals, but a different formulation of the task. Treating disease progression as a continuum, through regression on a continuous severity axis, ordinal modelling or longitudinal information, is in our view the natural next step and one to which the leakage-controlled, subject-level protocol established here can be directly extended.

\section*{Acknowledgements}

Data collection and sharing for the Alzheimer's Disease Neuroimaging Initiative (ADNI) is funded by the National Institute on Aging (National Institutes of Health Grant U19 AG024904). The grantee organization is the Northern California Institute for Research and Education. In the past, ADNI has also received funding from the National Institute of Biomedical Imaging and Bioengineering, the Canadian Institutes of Health Research, and private sector contributions through the Foundation for the National Institutes of Health (FNIH).

L. Cavinato is funded by the National Plan for NRRP Complementary Investments "Advanced Technologies for Human-centred Medicine" (PNC0000003). The present research is part of the activities of "Dipartimento di Eccellenza 2023-2027".

\appendix
\section{Appendix A}
\label{app:appendixA}
\subsection{PET-only training objective and checkpoint selection}
\label{app:pet_only_loss}

This appendix reports the training objective and validation criterion used for the PET-only Tau model described in Section~\ref{sec:methods}. The model is trained on scan-level Tau PET volumes, while validation and test predictions are aggregated at subject level by averaging the class probabilities of all scans belonging to the same subject.

\subsubsection{Class-balanced focal loss}

Let \(z_i \in \mathbb{R}^{K}\) denote the logits produced by the network for scan \(i\), with \(K=3\) classes corresponding to CN, MCI and AD. The predicted class probabilities are obtained by the softmax function,
\begin{equation}
    \hat{p}_{i,k}
    =
    \frac{\exp(z_{i,k})}
    {\sum_{j=1}^{K}\exp(z_{i,j})}.
\end{equation}

To reduce the effect of class imbalance, the PET-only network is trained with a class-balanced focal loss. Since label smoothing is used in the implementation, the loss is written in terms of the label-smoothed cross-entropy
\begin{equation}
    \ell_i^{\mathrm{LS}}
    =
    -\sum_{k=1}^{K}
    q_{i,k}^{\mathrm{LS}}
    \log \hat{p}_{i,k},
\end{equation}
where \(q_{i,k}^{\mathrm{LS}}\) is the label-smoothed target distribution. The focal term is then computed from
\begin{equation}
    p_{t,i}
    =
    \exp(-\ell_i^{\mathrm{LS}}),
\end{equation}
and the final training objective is
\begin{equation}
    \mathcal{L}_{\mathrm{CB\text{-}Focal}}
    =
    \frac{1}{N}
    \sum_{i=1}^{N}
    w_{y_i}\alpha_{y_i}
    (1-p_{t,i})^{\gamma}
    \ell_i^{\mathrm{LS}}.
    \label{eq:app_pet_cb_focal}
\end{equation}

The class-balanced weights are derived from the effective number of samples. For class \(k\), let \(n_k\) be the number of training samples in that class. The unnormalised class-balanced weight is
\begin{equation}
    \tilde{w}_k
    =
    \frac{1-\beta}{1-\beta^{n_k}},
\end{equation}
and the weights are normalised to have unit mean:
\begin{equation}
    w_k
    =
    \frac{\tilde{w}_k}
    {\frac{1}{K}\sum_{j=1}^{K}\tilde{w}_j}.
    \label{eq:app_pet_cb_weights}
\end{equation}

In the final configuration, the effective-number parameter is set to \(\beta=0.999\), the focal focusing parameter to \(\gamma=1.5\) and the label-smoothing coefficient to \(0.01\). In addition to the class-balanced weights \(w_k\), a mild manually specified class factor
\begin{equation}
    \alpha = [1.0,\;1.2,\;1.0]
\end{equation}
is used to slightly emphasize the MCI class.

\subsubsection{Subject-level probability aggregation}

Although the network is trained on scan-level PET volumes, all reported metrics are computed at subject level. For a subject \(s\) with scan set \(\mathcal{S}_s\), the subject-level probability vector is obtained by averaging the scan-level softmax outputs:
\begin{equation}
    \hat{\mathbf{p}}_s
    =
    \frac{1}{|\mathcal{S}_s|}
    \sum_{i\in\mathcal{S}_s}
    \hat{\mathbf{p}}_i.
    \label{eq:app_pet_subject_aggregation}
\end{equation}
The final subject-level PET-only prediction is then
\begin{equation}
    \hat{y}_s
    =
    \arg\max_{k\in\{\mathrm{CN},\mathrm{MCI},\mathrm{AD}\}}
    \hat{p}_{s,k}.
\end{equation}

\subsubsection{Checkpoint-selection score}

The best PET-only checkpoint is selected on the inner validation subjects using a cost-sensitive subject-level score. The score combines MCI F1, macro-F1 and balanced accuracy, while penalising clinically severe mistakes:
\begin{equation}
    S_{\mathrm{val}}
    =
    0.45\,F1_{\mathrm{MCI}}
    +
    0.35\,F1_{\mathrm{macro}}
    +
    0.20\,\mathrm{BalAcc}
    -
    0.10\,\frac{C}{N},
    \label{eq:app_pet_val_score}
\end{equation}
where \(N\) is the number of validation subjects and \(C\) is the error cost
\begin{equation}
    C
    =
    3\,FN_{\mathrm{AD}}
    +
    1\,FP_{\mathrm{AD}}
    +
    2\,N_{\mathrm{CN}\leftrightarrow\mathrm{AD}}
    +
    2\,FN_{\mathrm{CN}}.
    \label{eq:app_pet_cost}
\end{equation}

Here \(FN_{\mathrm{AD}}\) denotes AD subjects classified as either CN or MCI, \(FP_{\mathrm{AD}}\) denotes non-AD subjects classified as AD, \(N_{\mathrm{CN}\leftrightarrow\mathrm{AD}}\) denotes the number of direct CN--AD confusions and \(FN_{\mathrm{CN}}\) denotes CN subjects classified as either MCI or AD. This validation score is used only for checkpoint selection. The final PET-only predictions are still obtained by applying the argmax rule to the aggregated subject-level probabilities.

\subsubsection{Optimisation details}

The network is optimized with AdamW~\cite{loshchilov2019adamw}, using learning rate \(2\times 10^{-4}\) and weight decay \(10^{-2}\). Mixed-precision training, gradient accumulation and gradient clipping with maximum norm \(1.0\) are used to stabilize training. Early stopping is applied when the validation score in Equation~\eqref{eq:app_pet_val_score} does not improve for the specified patience window.

\subsection{Harvard--Oxford ROI feature set}
\label{app:roi-features}
For completeness, Table~\ref{tab:roi-features} lists the per-region features extracted from each Harvard--Oxford ROI on the normalised MRI and Tau PET volumes (Section~\ref{sec:dataset}). The same set of intensity summaries, robust $z$-scores, global-normalised ratios and quality-control indicators is computed for every cortical, subcortical and macro-region.

\begin{table}[H]
\centering
\caption{Per-region features extracted from each Harvard--Oxford ROI on the normalised volumes. The same set is computed for every cortical, subcortical and macro-region. Global statistics refer to the whole-brain intensity distribution of the same image.}
\label{tab:roi-features}
\begin{tabular}{lll}
\toprule
\textbf{Group} & \textbf{Feature} & \textbf{Description} \\
    \midrule
    \multirow{3}{*}{Intensity}
     & \texttt{mean}, \texttt{median}, \texttt{std}, \texttt{iqr} & Central tendency and dispersion \\
     & \texttt{p75}, \texttt{p90}, \texttt{p95} & Upper-tail percentiles \\
     & \texttt{tmean10} & $10\%$ trimmed mean \\
    \midrule
    \multirow{2}{*}{Robust $z$}
     & \texttt{zmean\_robust}, \texttt{zmedian\_robust} & Region statistic standardised \\
     & \texttt{zp90\_robust}, \texttt{ziqr\_robust} & against the global distribution \\
    \midrule
    \multirow{2}{*}{Ratios}
     & \texttt{median\_over\_global\_median} & \multirow{2}{*}{Regional-to-global intensity ratios} \\
     & \texttt{p90\_over\_global\_median}, \texttt{mean\_over\_global\_mean} & \\
    \midrule
    \multirow{2}{*}{QC}
     & \texttt{nvox}, \texttt{coverage} & Voxel count and fractional coverage \\
     & \texttt{used\_eroded} & Erosion fallback flag \\
    \bottomrule
\end{tabular}
\end{table}

\subsection{Tabular inner-model selection score}
\label{app:tabular_selection_score}

For the ROI/plasma tabular branch, model family, feature ablation, hyperparameters and ROI subset size are selected inside the outer-training fold by nested inner cross-validation. Candidate configurations are ranked using the following validation score:

\begin{equation}
\begin{aligned}
    S_{\mathrm{tab}}
    =\;&
    0.35\,F1_{\mathrm{macro}}
    + 0.25\,\mathrm{BalAcc}
    + 0.25\,F1_{\mathrm{MCI}}
    + 0.10\,F1_{\mathrm{AD}}
    + 0.05\,\mathrm{AUC}_{\mathrm{OvR,macro}}\\
    &- 0.002\,N_{\mathrm{CN}\rightarrow\mathrm{AD}}
    - 0.002\,N_{\mathrm{AD}\rightarrow\mathrm{CN}}.
\end{aligned}
\end{equation}

This score is used only for model selection inside the training data of each outer fold. The final reported metrics are computed exclusively on the held-out outer-fold subjects using the out-of-fold probabilities produced by the selected tabular ensemble.

\subsection{Hierarchical PET-plasma training details}
\label{app:hierarchical_details}

This appendix reports the stage-specific training losses and threshold-selection procedure used by the hierarchical PET-plasma branch.

\subsubsection{Stage-specific losses}

The hierarchical branch is composed of two binary stages. Stage 1 separates CN from non-CN subjects. The PET image model is trained as a two-class classifier using a focal loss on the two-class logits:
\begin{equation}
    \mathcal{L}_{\mathrm{Stage1}}
    =
    \frac{1}{N}
    \sum_{i=1}^{N}
    \alpha_{y_i}
    (1-p_{t,i})^{\gamma}
    \left(-\log p_{t,i}\right),
    \qquad
    \gamma=1.5,
\end{equation}
where \(p_{t,i}\) is the softmax probability assigned to the true binary class and \(\alpha_{y_i}\) is derived from the inverse class proportions and normalized to have unit mean.

Stage 2 separates MCI from AD subjects. The positive class is AD and the negative class is MCI. In this stage, the PET image model is trained with a convex combination of binary cross-entropy with logits and binary focal loss:
\begin{equation}
    \mathcal{L}_{\mathrm{Stage2}}
    =
    0.75\,\mathcal{L}_{\mathrm{BCE}}
    +
    0.25\,\mathcal{L}_{\mathrm{Focal}},
\end{equation}
with focal parameter \(\gamma=1.0\). The binary cross-entropy term uses a positive-class weight proportional to the ratio between the number of MCI and AD training samples.

\subsubsection{PET-plasma meta-model}

For each stage, the PET model produces a subject-level probability by averaging scan-level predictions across scans belonging to the same subject. A plasma-only logistic model is trained on the available plasma biomarkers. The PET and plasma outputs are then combined by a logistic meta-model using probability-, uncertainty- and agreement-based features derived from the two sources. The plasma-only model median-imputes missing markers inside its training-fitted pipeline. In the meta-model, by contrast, a subject without a plasma prediction receives a neutral plasma score ($\hat{p}_{\text{plasma}} = 0.5$, with zero margin and logit) rather than an imputed biomarker value, and the availability of plasma is passed explicitly through per-marker missingness flags, the availability counts $\texttt{has\_plasma\_all3}$, $\texttt{has\_plasma\_any}$ and $\texttt{n\_plasma\_available}$, and interaction terms between the branch scores and the availability flag. This lets the meta-model down-weight the plasma channel for the subjects in which it is absent, so that those subjects contribute through the PET signal rather than through a fabricated biomarker value.

\subsubsection{Hierarchical probability recomposition and final decision}

Let \(\hat{p}_1\) denote the meta-model probability of being non-CN and let \(\hat{p}_2\) denote the meta-model probability of AD. The multiclass probability vector is computed as
\begin{equation}
    P(\mathrm{CN}) = 1-\hat{p}_1,
    \qquad
    P(\mathrm{MCI}) = \hat{p}_1(1-\hat{p}_2),
    \qquad
    P(\mathrm{AD}) = \hat{p}_1\hat{p}_2.
\end{equation}

The final hard decision is not obtained by a simple argmax over these probabilities. Instead, validation-tuned stage thresholds are used. A subject is assigned to CN if
\begin{equation}
    \hat{p}_1 < \tau_1.
\end{equation}
Among the remaining subjects, AD is assigned only if
\begin{equation}
    \hat{p}_2 \geq \tau_2
    \quad \text{and} \quad
    \hat{p}_1 \geq 0.45.
\end{equation}
Otherwise, the subject is assigned to MCI.

\subsubsection{Threshold-selection score}

The thresholds \(\tau_1\) and \(\tau_2\) are selected only on the inner-validation subjects of each outer fold. Candidate thresholds are ranked using a multiclass validation score combining balanced accuracy, macro-F1, MCI F1 and a penalty against direct CN--AD confusions:
\begin{equation}
    S_{\mathrm{HIER}}
    =
    0.30\,\mathrm{BalAcc}
    +
    0.25\,F1_{\mathrm{macro}}
    +
    0.20\,F1_{\mathrm{MCI}}
    +
    0.25\,(1-P_{\mathrm{CN\leftrightarrow AD}}),
\end{equation}
where
\begin{equation}
    P_{\mathrm{CN\leftrightarrow AD}}
    =
    \frac{N_{\mathrm{CN}\rightarrow\mathrm{AD}} + N_{\mathrm{AD}\rightarrow\mathrm{CN}}}{N}.
\end{equation}
This score is used only for threshold selection. Final evaluation is performed once on the held-out outer-fold subjects.

\subsection{Multimodal superlearner fusion details}
\label{app:superlearner_details}

This appendix details the objective and the weight estimation of the nested
convex superlearner used as the robustness analysis of the fixed convex fusion
(Section~\ref{subsec:fusion}). Throughout, $\mathbf{P} \in \mathbb{R}^{N \times 4 \times 3}$
denotes the tensor of out-of-fold branch probabilities ($N$ subjects, $4$
branches, $3$ classes) and $y_s \in \{0,1,2\}$ the true label of subject $s$.

\subsubsection{Class-balanced negative log-likelihood}

The fusion weights are estimated by minimizing a class-balanced negative
log-likelihood of the fused probabilities. Each subject is reweighted by the
inverse frequency of its class, normalized to unit mean, so that the rare AD and
MCI subjects are not dominated by the majority CN class. Let
$n_c = \sum_s \mathbb{1}[y_s = c]$ be the class counts and
\begin{equation}
    \omega_s
    =
    \frac{1/n_{y_s}}{\frac{1}{N}\sum_{s'} 1/n_{y_{s'}}}
\end{equation}
the normalized per-subject weight. The class-balanced negative log-likelihood of
a fused probability matrix $\bar{\mathbf{p}}$ is
\begin{equation}
    \mathcal{L}_{\text{CB-NLL}}(y, \bar{\mathbf{p}})
    =
    \frac{1}{N}
    \sum_{s=1}^{N}
    \omega_s \,\big(-\log \bar{p}_{s,y_s}\big).
    \label{eq:app_cbnll}
\end{equation}

\subsubsection{Global and classwise convex fusion}

Two fusion families are considered. In the global family, a single weight
vector $\mathbf{w} \in \mathbb{R}^4$ is shared across classes and the fused
probability of subject $s$ is the convex combination of the branch probabilities,
\begin{equation}
    \bar{\mathbf{p}}_s
    =
    \mathcal{N}\!\Big(\sum_{m=1}^{4} w_m \, \mathbf{p}^{(m)}_s\Big),
    \qquad
    \sum_m w_m = 1, \; w_m \ge 0,
\end{equation}
where $\mathcal{N}(\cdot)$ renormalizes the vector to sum to one. In the
classwise family, each class $c$ is assigned its own convex weight vector
$\mathbf{W}_{c,:} \in \mathbb{R}^4$ and the fusion is performed per class before
renormalization,
\begin{equation}
    \bar{p}_{s,c}
    \;\propto\;
    \sum_{m=1}^{4} W_{c,m}\, p^{(m)}_{s,c},
    \qquad
    \sum_m W_{c,m} = 1, \; W_{c,m} \ge 0,
\end{equation}
which allows, for example, the MCI class to draw more heavily on the branches
that are most informative for it while the CN class relies on others.

\subsubsection{Softmax parametrization and anchored objective}

To enforce the non-negativity and sum-to-one constraints without constrained
optimization, the weights are parametrized as a softmax of unconstrained logits
$\mathbf{z}$, i.e.\ $\mathbf{w} = \mathrm{softmax}(\mathbf{z})$ (and row-wise for
the classwise matrix $\mathbf{W}$). The logits are estimated by minimizing the
class-balanced NLL of Equation~\eqref{eq:app_cbnll} augmented with a quadratic
penalty that anchors the weights towards a reference anchor $\mathbf{a}$
(itself normalized to sum to one),
\begin{equation}
    \min_{\mathbf{z}}
    \;\;
    \mathcal{L}_{\text{CB-NLL}}\big(y, \bar{\mathbf{p}}(\mathbf{z})\big)
    \;+\;
    \lambda \, \big\lVert \mathbf{w}(\mathbf{z}) - \mathbf{a} \big\rVert_2^2,
    \label{eq:app_anchored}
\end{equation}
with $\lambda \ge 0$. The optimization is solved by quasi-Newton (BFGS) minimization, initialized at $\mathbf{z}_0 = \log \mathbf{a}$, so that at $\lambda \to \infty$ the solution collapses to the fixed anchor and at $\lambda = 0$ it is the unconstrained class-balanced maximum-likelihood combination. The penalty therefore interpolates smoothly between the learned and the fixed fusion and makes explicit how far the data-driven optimum departs from the a-priori anchor.

\subsubsection{Nested selection and anchor sensitivity}

For each anchor $\mathbf{a}$, the fusion family
(global or classwise) and the regularization strength
$\lambda \in \{0, 0.01, 0.03, 0.1, 0.3, 1, 3, 10\}$ are selected by an inner
cross-validation over the synchronized folds, restricted to the outer-training
subjects. Candidate configurations are ranked by a class-balanced selection
score
\begin{equation}
    \begin{aligned}
    S_{\text{SL}}
    =\;&
    0.23\,\mathrm{Acc}
    + 0.23\,F1_{\mathrm{macro}}
    + 0.18\,\mathrm{BalAcc}
    + 0.12\,F1_{\mathrm{MCI}}
    + 0.10\,F1_{\mathrm{AD}} \\
    &+ 0.07\,\mathrm{AUC}_{\mathrm{OvR,macro}}
    + 0.07\,\mathrm{AP}_{\mathrm{macro}}
    - 0.0045\,N_{\mathrm{CN}\leftrightarrow\mathrm{AD}},
    \end{aligned}
    \label{eq:app_sl_score}
\end{equation}
where $N_{\mathrm{CN}\leftrightarrow\mathrm{AD}}$ is the number of severe
CN--AD confusions. The selected configuration is refitted on the outer-training
folds and applied once to the held-out outer fold. The whole procedure is
repeated over the three anchors (uniform, mild-tabular and the primary
tabular anchor) to assess the sensitivity of the learned fusion to the anchor
choice. The same selection score of Equation~\eqref{eq:app_sl_score} and the
same nested protocol are used for the logistic meta-learner, whose inverse
regularization strength $C \in \{0.03, 0.1, 0.3, 1, 3, 10\}$ and class weights
are selected on the inner folds, without posterior probability calibration.

\subsection{Calibration ablation of the logistic meta-learner}
\label{app:calibration_ablation}

The logistic meta-learner is reported without any posterior probability
calibration of its outputs. This appendix motivates that choice through an
ablation in which the meta-learner probabilities are post-hoc calibrated, on the
inner-validation subjects only, using either Platt scaling (sigmoid) or isotonic
regression~\cite{guo2017calibration} and compared against the uncalibrated
variant. All three variants share the same nested selection protocol and differ
only in the calibration step; metrics are computed on the pooled out-of-fold
subject-level predictions.

\begin{table}[H]
    \centering
    \caption{Effect of post-hoc calibration on the logistic meta-learner. Both
    calibration methods improve overall accuracy but collapse the MCI class,
    halving its recall, whereas the uncalibrated variant attains the best MCI
    behaviour and the lowest calibration error.} 
    \label{tab:calib-ablation}
    \begin{tabular}{lccccc}
    \toprule
    \textbf{Calibration} & \textbf{Acc} & \textbf{F1$_{\mathrm{MCI}}$} & \textbf{Recall$_{\mathrm{MCI}}$} & \textbf{ECE} & \textbf{CN$\leftrightarrow$AD} \\
    \midrule
    None (used)      & $0.687$ & $\mathbf{0.482}$ & $\mathbf{0.456}$ & $\mathbf{0.032}$ & $\mathbf{9}$ \\
    Sigmoid (Platt)  & $\mathbf{0.700}$ & $0.328$ & $0.221$ & $0.065$ & $18$ \\
    Isotonic         & $0.696$ & $0.306$ & $0.202$ & $0.045$ & $19$ \\
    \bottomrule
    \end{tabular}
\end{table}

The ablation shows a clear trade-off. Both calibration methods marginally
increase overall accuracy, by sharpening the decision in favour of the majority
CN class, but in doing so they roughly halve the MCI recall (from $0.456$ to
$0.221$ and $0.202$) and approximately double the number of severe
CN$\leftrightarrow$AD confusions. The reason is structural: calibration is fitted
on the small inner-validation folds, where the rare MCI and AD classes are poorly
represented, so the calibrated mapping is dominated by the majority class and
suppresses the minority-class probabilities that the meta-learner needs in order
to recover MCI. Notably, the uncalibrated meta-learner already attains the lowest
expected calibration error ($\text{ECE}=0.032$), indicating that its raw outputs
are well calibrated without any post-hoc correction. Since the central objective
of this work is the recovery of the MCI class and since calibration degrades
exactly that objective while providing no calibration benefit, the uncalibrated
meta-learner is retained as the reference fusion model.

\section{Appendix B}
\label{app:appendixB}

This appendix collects the complete pooled and fold-level metric panels for the
four branches and the multimodal fusion, whose confusion matrices and headline
metrics are reported in Section~\ref{sec:results}. For every model, pooled
out-of-fold (OOF) metrics are computed from all $881$ subject-level predictions,
and fold values are reported as mean $\pm$ standard deviation across the five
outer folds. The branch subsections follow the same order as Section~\ref{sec:results}.

\subsection{Confusion-matrix grid}
\label{app:cm_grid}

\begin{figure}[H]
    \centering
    \includegraphics[width=\textwidth]{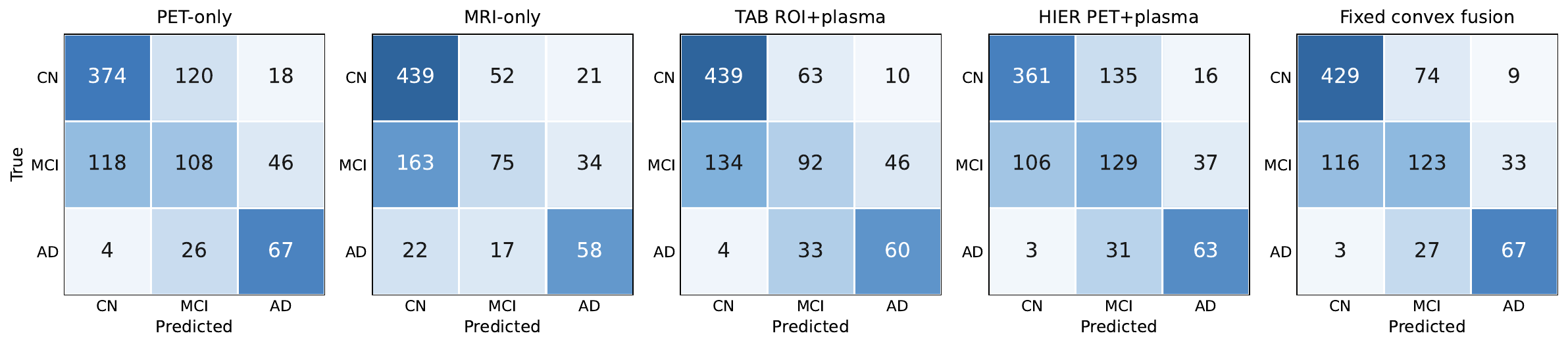}
    \caption{Pooled subject-level confusion matrices of the four branches and
    the fixed convex fusion on the synchronized $881$-subject cohort. Rows
    correspond to true classes and columns to predicted classes; cell colour
    encodes the row-normalised proportion (per true class) and the text reports
    the subject counts. Across all models the dominant off-diagonal mass is the
    MCI$\rightarrow$CN confusion, while severe CN$\leftrightarrow$AD confusions
    remain rare.}
    \label{fig:cm-grid}
\end{figure}

Figure~\ref{fig:cm-grid} shows the five pooled confusion matrices side by side,
with cell colour encoding the row-normalised (per-true-class) proportion and
cell text the subject counts.

\subsection{PET-only branch}
\label{app:full_metrics_pet}

\begin{table}[H]
    \centering
    \caption{Performance of the PET-only branch on the synchronized cohort. Pooled OOF metrics are computed from all $881$ subject-level predictions. Fold values are reported as mean $\pm$ standard deviation across the five outer folds.}
    \label{tab:pet-overall-app}
    \begin{tabular}{lcc}
    \toprule
    \textbf{Metric} & \textbf{Pooled OOF} & \textbf{Fold mean $\pm$ std} \\
    \midrule
    Accuracy & $0.623$ & $0.623 \pm 0.047$ \\
    Balanced accuracy & $0.606$ & $0.607 \pm 0.045$ \\
    Macro-F1 & $0.580$ & $0.579 \pm 0.026$ \\
    Weighted-F1 & $0.623$ & $0.621 \pm 0.040$ \\
    \midrule
    F1$_{\mathrm{CN}}$ & $0.742$ & $0.739 \pm 0.064$ \\
    F1$_{\mathrm{MCI}}$ & $0.411$ & $0.410 \pm 0.042$ \\
    F1$_{\mathrm{AD}}$ & $0.588$ & $0.587 \pm 0.038$ \\
    \midrule
    Recall$_{\mathrm{CN}}$ & $0.730$ & $0.731 \pm 0.110$ \\
    Recall$_{\mathrm{MCI}}$ & $0.397$ & $0.397 \pm 0.065$ \\
    Recall$_{\mathrm{AD}}$ & $0.691$ & $0.693 \pm 0.123$ \\
    \midrule
    AUC (macro OvR) & $0.746$ & $0.749 \pm 0.035$ \\
    AP (macro OvR) & $0.573$ & $0.593 \pm 0.048$ \\
    Log-loss & $0.802$ & $0.802 \pm 0.047$ \\
    ECE (10 bins) & $0.075$ & --- \\
    \bottomrule
    \end{tabular}
\end{table}

\subsection{Structural MRI branch}
\label{app:full_metrics_mri}

\begin{table}[H]
    \centering
    \caption{Performance of the structural MRI branch on the synchronized cohort. Pooled OOF metrics are computed from all $881$ subject-level predictions. Fold values are reported as mean $\pm$ standard deviation across the five outer folds.}
    \label{tab:mri-overall-app}
    \begin{tabular}{lcc}
    \toprule
    \textbf{Metric} & \textbf{Pooled OOF} & \textbf{Fold mean $\pm$ std} \\
    \midrule
    Accuracy & $0.649$ & $0.649 \pm 0.047$ \\
    Balanced accuracy & $0.577$ & $0.578 \pm 0.043$ \\
    Macro-F1 & $0.562$ & $0.557 \pm 0.058$ \\
    Weighted-F1 & $0.621$ & $0.613 \pm 0.039$ \\
    \midrule
    F1$_{\mathrm{CN}}$ & $0.773$ & $0.768 \pm 0.053$ \\
    F1$_{\mathrm{MCI}}$ & $0.361$ & $0.341 \pm 0.115$ \\
    F1$_{\mathrm{AD}}$ & $0.552$ & $0.562 \pm 0.110$ \\
    \midrule
    Recall$_{\mathrm{CN}}$ & $0.857$ & $0.857 \pm 0.147$ \\
    Recall$_{\mathrm{MCI}}$ & $0.276$ & $0.276 \pm 0.161$ \\
    Recall$_{\mathrm{AD}}$ & $0.598$ & $0.600 \pm 0.117$ \\
    \midrule
    AUC (macro OvR) & $0.718$ & $0.729 \pm 0.057$ \\
    AP (macro OvR) & $0.546$ & $0.573 \pm 0.068$ \\
    Log-loss & $0.813$ & $0.813 \pm 0.091$ \\
    ECE (10 bins) & $0.060$ & --- \\
    \bottomrule
    \end{tabular}
\end{table}

\subsection{Tabular ROI-plasma branch}
\label{app:full_metrics_tab}

\begin{table}[H]
    \centering
    \caption{Performance of the tabular ROI--plasma branch on the synchronized cohort. Pooled OOF metrics are computed from all $881$ subject-level predictions. Fold values are reported as mean $\pm$ standard deviation across the five outer folds.}
    \label{tab:tab-overall-app}
    \begin{tabular}{lcc}
    \toprule
    \textbf{Metric} & \textbf{Pooled OOF} & \textbf{Fold mean $\pm$ std} \\
    \midrule
    Accuracy & $0.671$ & $0.671 \pm 0.016$ \\
    Balanced accuracy & $0.605$ & $0.606 \pm 0.052$ \\
    Macro-F1 & $0.590$ & $0.588 \pm 0.032$ \\
    Weighted-F1 & $0.654$ & $0.653 \pm 0.017$ \\
    \midrule
    F1$_{\mathrm{CN}}$ & $0.806$ & $0.806 \pm 0.012$ \\
    F1$_{\mathrm{MCI}}$ & $0.400$ & $0.400 \pm 0.025$ \\
    F1$_{\mathrm{AD}}$ & $0.563$ & $0.557 \pm 0.081$ \\
    \midrule
    Recall$_{\mathrm{CN}}$ & $0.857$ & $0.858 \pm 0.031$ \\
    Recall$_{\mathrm{MCI}}$ & $0.338$ & $0.338 \pm 0.032$ \\
    Recall$_{\mathrm{AD}}$ & $0.619$ & $0.622 \pm 0.170$ \\
    \midrule
    AUC (macro OvR) & $0.791$ & $0.794 \pm 0.020$ \\
    AP (macro OvR) & $0.603$ & $0.625 \pm 0.027$ \\
    Log-loss & $0.714$ & $0.714 \pm 0.029$ \\
    ECE (10 bins) & $0.054$ & --- \\
    \bottomrule
    \end{tabular}
\end{table}

\begin{table}[H]
    \centering
    \caption{Selection frequency of Harvard--Oxford ROI groups across the five
    outer folds for the tabular ROI--plasma branch. A region is counted as selected in a fold if
    at least one of its intensity summaries enters the retained top-$k$ subset.
    Only regions selected in at least two folds are listed.}
    \label{tab:roi-selection}
    \begin{tabular}{lc}
    \toprule
    \textbf{Region (Harvard--Oxford / macro)} & \textbf{Folds (/5)} \\
    \midrule
    Tau AD-signature (macro)              & $5$ \\
    Amyloid composite (macro)             & $5$ \\
    Medial temporal lobe (macro)          & $5$ \\
    Lateral temporal (macro)              & $5$ \\
    Posterior default-mode (macro)        & $5$ \\
    Parietal (macro)                      & $5$ \\
    Left amygdala                         & $5$ \\
    Right amygdala                        & $5$ \\
    Precuneus                             & $5$ \\
    Posterior cingulate gyrus             & $5$ \\
    Angular gyrus                         & $5$ \\
    Middle temporal gyrus (post.\ div.)   & $5$ \\
    Middle temporal gyrus (temp.-occ.)    & $5$ \\
    \midrule
    Left hippocampus                      & $3$ \\
    Superior temporal gyrus (post.\ div.) & $2$ \\
    Inferior temporal gyrus (temp.-occ.)  & $2$ \\
    Parietal operculum                    & $2$ \\
    Cingulate (macro)                     & $2$ \\
    \bottomrule
    \end{tabular}
\end{table}

\subsection{Hierarchical PET--plasma branch}
\label{app:full_metrics_hier}

\begin{table}[H]
\centering
\caption{Performance of the hierarchical PET--plasma branch on the synchronized cohort. Pooled OOF metrics are computed from all $881$ subject-level predictions. Fold values are reported as mean $\pm$ standard deviation across the five outer folds.}
\label{tab:hier-overall-app}
\begin{tabular}{lcc}
\toprule
\textbf{Metric} & \textbf{Pooled OOF} & \textbf{Fold mean $\pm$ std} \\
    \midrule
    Accuracy & $0.628$ & $0.628 \pm 0.092$ \\
    Balanced accuracy & $0.610$ & $0.610 \pm 0.071$ \\
    Macro-F1 & $0.594$ & $0.588 \pm 0.087$ \\
    Weighted-F1 & $0.633$ & $0.625 \pm 0.088$ \\
    \midrule
    F1$_{\mathrm{CN}}$ & $0.735$ & $0.728 \pm 0.095$ \\
    F1$_{\mathrm{MCI}}$ & $0.455$ & $0.443 \pm 0.141$ \\
    F1$_{\mathrm{AD}}$ & $0.592$ & $0.594 \pm 0.099$ \\
    \midrule
    Recall$_{\mathrm{CN}}$ & $0.705$ & $0.705 \pm 0.166$ \\
    Recall$_{\mathrm{MCI}}$ & $0.474$ & $0.473 \pm 0.181$ \\
    Recall$_{\mathrm{AD}}$ & $0.649$ & $0.653 \pm 0.178$ \\
    \midrule
    AUC (macro OvR) & $0.760$ & $0.769 \pm 0.038$ \\
    AP (macro OvR) & $0.576$ & $0.606 \pm 0.058$ \\
    Log-loss & $0.754$ & $0.754 \pm 0.058$ \\
    ECE (10 bins) & $0.040$ & --- \\
    \bottomrule
\end{tabular}
\end{table}

\begin{table}[H]
\centering
\caption{Stage-wise binary performance of the hierarchical PET--plasma branch, reported as mean across the five outer folds. For Stage~1, the positive class is non-CN. For Stage~2, the positive class is AD and all outer-test subjects are included before final hierarchical recomposition.}
\label{tab:hier-stages-app}
\begin{tabular}{llccccc}
\toprule
\textbf{Stage} & \textbf{Source} & \textbf{Acc} & \textbf{Bal.\ Acc} & \textbf{F1} & \textbf{Recall} & \textbf{AUC} \\
\midrule
\multirow{4}{*}{\shortstack[l]{Stage 1\\ CN vs non-CN}}
& PET-only             & $0.715$ & $0.689$ & $0.597$ & $0.526$ & $0.761$ \\
& Plasma-only          & $0.693$ & $0.663$ & $0.564$ & $0.474$ & $0.710$ \\
& Average PET$+$plasma & $0.730$ & $0.701$ & $0.614$ & $0.523$ & $0.773$ \\
& Meta PET$+$plasma    & $0.738$ & $\mathbf{0.719}$ & $0.653$ & $0.599$ & $0.771$ \\
\midrule
\multirow{4}{*}{\shortstack[l]{Stage 2\\ AD vs non-AD}}
& PET-only             & $0.910$ & $0.693$ & $0.494$ & $0.414$ & $0.906$ \\
& Plasma-only          & $0.869$ & $0.733$ & $0.485$ & $0.558$ & $0.831$ \\
& Average PET$+$plasma & $0.915$ & $0.727$ & $0.549$ & $0.486$ & $0.900$ \\
& Meta PET$+$plasma    & $0.889$ & $\mathbf{0.812}$ & $0.591$ & $0.714$ & $0.905$ \\
\bottomrule
\end{tabular}
\end{table}

\subsection{Fixed convex fusion}
\label{app:full_metrics_fusion}

\begin{table}[H]
    \centering
    \caption{Performance of the fixed convex fusion (primary multimodal model) on the synchronized cohort. Pooled OOF metrics are computed from all $881$ subject-level predictions; fold values are reported as mean $\pm$ standard deviation across the five outer folds.}
    \label{tab:fusion-overall-app}
    \begin{tabular}{lcc}
    \toprule
    \textbf{Metric} & \textbf{Pooled OOF} & \textbf{Fold mean $\pm$ std} \\
    \midrule
    Accuracy            & $0.703$ & $0.703 \pm 0.039$ \\
    Balanced accuracy   & $0.660$ & $0.661 \pm 0.063$ \\
    Macro-F1            & $0.652$ & $0.650 \pm 0.048$ \\
    Weighted-F1         & $0.695$ & $0.694 \pm 0.039$ \\
    \midrule
    F1$_{\mathrm{CN}}$  & $0.809$ & $0.809 \pm 0.032$ \\
    F1$_{\mathrm{MCI}}$ & $0.496$ & $0.495 \pm 0.058$ \\
    F1$_{\mathrm{AD}}$  & $0.646$ & $0.646 \pm 0.076$ \\
    \midrule
    Recall$_{\mathrm{CN}}$  & $0.838$ & $0.838 \pm 0.057$ \\
    Recall$_{\mathrm{MCI}}$ & $0.452$ & $0.452 \pm 0.066$ \\
    Recall$_{\mathrm{AD}}$  & $0.691$ & $0.694 \pm 0.168$ \\
    \midrule
    AUC (macro OvR)     & $0.804$ & $0.804 \pm 0.030$ \\
    AP (macro OvR)      & $0.635$ & $0.660 \pm 0.050$ \\
    Log-loss            & $0.710$ & $0.710 \pm 0.041$ \\
    ECE (10 bins)       & $0.094$ & --- \\
    \bottomrule
    \end{tabular}
\end{table}

\subsection{Logistic meta-learner}
\label{app:full_metrics_logistic_meta}

\begin{table}[H]
\centering
\caption{Performance of the leakage-controlled logistic meta-learner on the synchronized cohort. Pooled OOF metrics are computed from all $881$ subject-level predictions. Fold values are reported as mean $\pm$ standard deviation across the five outer folds.}
\label{tab:logistic-meta-overall-app}
\begin{tabular}{lcc}
    \toprule
    \textbf{Metric} & \textbf{Pooled OOF} & \textbf{Fold mean $\pm$ std} \\
    \midrule
    Accuracy & $0.672$ & $0.672 \pm 0.032$ \\
    Balanced accuracy & $0.614$ & $0.615 \pm 0.043$ \\
    Macro-F1 & $0.630$ & $0.628 \pm 0.036$ \\
    Weighted-F1 & $0.675$ & $0.673 \pm 0.027$ \\
    \midrule
    F1$_{\mathrm{CN}}$ & $0.776$ & $0.775 \pm 0.045$ \\
    F1$_{\mathrm{MCI}}$ & $0.507$ & $0.506 \pm 0.028$ \\
    F1$_{\mathrm{AD}}$ & $0.608$ & $0.604 \pm 0.096$ \\
    \midrule
    Recall$_{\mathrm{CN}}$ & $0.770$ & $0.770 \pm 0.084$ \\
    Recall$_{\mathrm{MCI}}$ & $0.537$ & $0.537 \pm 0.061$ \\
    Recall$_{\mathrm{AD}}$ & $0.536$ & $0.539 \pm 0.137$ \\
    \midrule
    AUC (macro OvR) & $0.803$ & $0.810 \pm 0.028$ \\
    AP (macro OvR) & $0.653$ & $0.685 \pm 0.048$ \\
    Log-loss & $0.704$ & $0.704 \pm 0.048$ \\
    \bottomrule
\end{tabular}
\end{table}

\begin{table}[H]
\centering
\caption{Pooled subject-level confusion matrix of the logistic meta-learner. Rows correspond to true classes and columns to predicted classes.}
\label{tab:logistic-meta-cm}
\begin{tabular}{lccc}
    \toprule
    & $\widehat{\mathrm{CN}}$ & $\widehat{\mathrm{MCI}}$ & $\widehat{\mathrm{AD}}$ \\
    \midrule
    \textbf{CN}  & $394$ & $114$ & $4$ \\
    \textbf{MCI} & $108$ & $146$ & $18$ \\
    \textbf{AD}  & $1$   & $44$  & $52$ \\
    \bottomrule
\end{tabular}
\end{table}

\end{document}